\documentclass[twocolumn]{aastex631}

\usepackage{graphicx}
\usepackage{epsfig}
\usepackage{gensymb}
\usepackage{amsmath}	
\usepackage{amssymb}	
\usepackage{verbatim}
\usepackage{color}
\usepackage[caption=false]{subfig}

\shorttitle{Refractive Index \& Extinction Coefficient of Water Ice}
\shortauthors{He et al.}

\submitjournal{ApJ}

\begin{document}

\title{Refractive index and extinction coefficient of vapor-deposited water ice in the UV-Vis range}

\correspondingauthor{Jiao He}
\email{he@mpia.de}

\author[0000-0003-2382-083X]{Jiao He}
\affiliation{Laboratory for Astrophysics, Leiden Observatory, Leiden University, P.O. Box 9513, NL 2300 RA Leiden, The Netherlands}
\affiliation{Current address: Max Planck Institute for Astronomy, K{\"o}nigstuhl 17, D-69117 Heidelberg, Germany}

\author{Sharon J. M. Diamant}
\affiliation{Laboratory for Astrophysics, Leiden Observatory, Leiden University, P.O. Box 9513, NL 2300 RA Leiden, The Netherlands}

\author{Siyu Wang}
\affiliation{Laboratory for Astrophysics, Leiden Observatory, Leiden University, P.O. Box 9513, NL 2300 RA Leiden, The Netherlands}

\author[0000-0002-0971-6078]{Haochuan Yu}
\affiliation{Laboratory for Astrophysics, Leiden Observatory, Leiden University, P.O. Box 9513, NL 2300 RA Leiden, The Netherlands}
\affiliation{Department of Astronomy, Beijing Normal University, Beijing 100875, China}

\author[0000-0001-6144-4113]{Will R. M. Rocha}
\affiliation{Laboratory for Astrophysics, Leiden Observatory, Leiden University, P.O. Box 9513, NL 2300 RA Leiden, The Netherlands}

\author{Marina Rachid}
\affiliation{Laboratory for Astrophysics, Leiden Observatory, Leiden University, P.O. Box 9513, NL 2300 RA Leiden, The Netherlands}

\author{Harold Linnartz}
\affiliation{Laboratory for Astrophysics, Leiden Observatory, Leiden University, P.O. Box 9513, NL 2300 RA Leiden, The Netherlands}

\begin{abstract}
Laboratory results of the optical properties of vapor-deposited water ice, specifically the refractive index and extinction coefficient, are available mainly for a selective set of wavelengths and a limited number of deposition temperatures. Experimental limitations are the main reason for the lack of broadband data, which is unfortunate as these quantities are needed to interpret and predict astronomical and planetary observations. 
The goal of this work is to address these lacking data, using an experimental broadband method that is capable of rapidly providing reliable water ice data across the entire UV-visible range.
This approach combines the simultaneous use of a monochromatic HeNe laser and a broadband Xe-arc lamp to record interference fringes of water ice during deposition at astronomically relevant ice temperatures. The ice thickness is typically more than 20 $\mu$m. Analyzing the period and intensity patterns combining both the monochromatic and broadband interference patterns allows the determination of the wavelength-dependent refractive index and extinction coefficient.
We present accurate refractive index and extinction coefficient graphs for wavelengths between 250 and 750 nm and ices deposited between 30 and 160 K. From our data, we find a possible structural change in the ice in the 110-130 K region that has not been reported before. We also discuss that the data presented in this paper can be used to interpret astronomical observations of icy surfaces.

\end{abstract}

\section{Introduction}\label{sec:intro}
Vapor-deposited water ice has attracted much attention in astrophysics and planetary sciences research because of its relevance in many space environments, such as the interstellar medium \citep[ISM, e.g.,][]{Boogert2015, vanDishoeck2021} and the surface of comets or icy moons in the Solar System \citep[e.g.,][]{Encrenaz2008, Mahieux2019, Spilker2019}. In ISM studies, the knowledge of the optical properties and structure of water ice is important to interpret astronomical observations of Young Stellar Objects (YSOs) in star-forming regions. Direct comparison of water ice laboratory spectra with the IR data of YSOs indicates that the majority of water ice is in the amorphous phase \citep[e.g.,][]{Boogert2008, Perotti2020}, although also higher degrees of crystallinity have been observed toward high-mass protostars \citep{Dartois2002}.

In Solar System environments, analysis of the 67P/Churyumov-Gerasimenko surface hints at the existence of crystalline water ice \citep[][]{Filacchione2016} of millimeter sizes \citep[][]{Barucci2016}. Similarly, spectral analysis of icy moons around Saturn (e.g., Enceladus, Iapetus, Rhea) and Jupiter (Europa, Ganymede, Callisto) has pointed to the presence of water-rich environments with different populations of grain sizes \citep[e.g.,][]{Iess2014, ROBIDEL2020, Paganini2020}. Data from the Galileo spacecraft obtained with the {\it Near-Infrared Mapping Spectrometer} (NIMS) showed evidence of crystalline water ice distributed on the surface of icy moons around Jupiter \citep[e.g.,][]{Grasset2017}. This is rather expected due to the interaction with the Jovian magnetosphere. Amorphous water ice was observed at the poles of Ganymede, which are protected from impacting magnetospheric particles \citep{Hansen2000b}. On Europa, \citet{Hansen2000a} showed that neither crystalline nor amorphous ices fit well the NIMS spectrum, although the low temperature would favor the presence of amorphous water ice. On the surfaces of the icy moons around Saturn, data from the {\it Visual and Infrared Mapping Spectrometer} (VIMS) onboard the Cassini probe, indicate that there is no amorphous ice in the Saturn system \citep{Filacchione2010, Clark2012, Clark2013, Clark2014, Buratti2018, Postberg2018}. 
On Enceladus, crystalline ice has been found close to the so-called ``tiger-strip'' cracks \citep{DHINGRA2017}. On Iapetus, a bimodal color is observed in the VIMS images, where the brightest region is dominated by water ice \citep{Clark2012}. On Rhea, the crystalline ice is mostly observed in the center of the craters \citep{dalle2015}. Throughout the Solar System, observations evaluated by \citet{DalleOre2018}, \citet{Cook2019}  and \citet{Cruikshank2020} concluded that crystalline ice dominates the spectra signatures of icy satellites in the outer Solar System with no evidence of amorphous ice, except in the cases mentioned previously for Europa and Ganymede in the inner Jupiter system.

Given the ubiquitous presence of crystalline water in the outer Solar System, and amorphous ice in the ISM, it is clear that there is the need for accurate laboratory data of fundamental properties of water ice - specifically refractive indices and extinction coefficients - as a function of wavelength and in dependence of varying parameters, such as temperature. Ultimately, these data are required to produce reliable synthetic spectra to be compared with astronomical observations \citep{Boogert2000, hapke2008, Clark2012, hapke_2012_book, Pontoppidan2005, rocha2015, Merlin2021}.

Previous laboratory studies have shown that the structure of vapor-deposited water ice is determined to a large extent by the substrate temperature and the incident angle at which the vapor is deposited \citep[e.g.,][]{Stevenson1999}. In general, at deposition temperatures lower than $\sim$130 K, the ice has an amorphous structure \citep{Fletcher1971} and therefore is referred to as amorphous solid water (ASW). At deposition temperatures above 160 K, the ice is usually crystalline. Between 130 and 160 K, additional factors may play a role, e.g., the rate of the vapor deposition and the time that the ice resides in a specific temperature range. In general, higher deposition temperature facilitates the amorphous--crystalline phase transition. Prior studies have also shown that high energy particles bombardment makes crystalline water ice amorphous \citep[see e.g.][]{Moore1992, Fama2010, Dartois2015}. Here in this study we only focus on ``thermal'' processes, and do not consider the ``energetic processes'' induced by high energy particles. Under pure thermal processes, the transition from amorphous ice to a  crystalline form is irreversible, meaning that the detection of crystalline ice indicates that the ice was once amorphous, at some point the temperature of the ice has increased. However, in the context of planetary surfaces (icy moons, comets, asteroids), a high deposition rate of vapor water onto a regolith could also lead to the formation of crystalline water even at low temperatures. In this study, we present the optical properties, including the refractive index and extinction coefficient of water ice, both amorphous and crystalline,  deposited at different temperatures and covering the UV-Vis range roughly from 250 to 750 nm.

In the literature, laboratory measurements of the refractive index of ASW have been reported by \citet{Berland1995, Brown1996, Westley1998, Dohnalek2003}. These prior studies measured Helium-Neon (HeNe) laser interference during water vapor deposition at different temperatures and calculated the refractive index of water ice for one wavelength, the wavelength at which the HeNe laser emits, 632.8 nm. Typically a few interference fringes have been measured, hence limiting the accuracy of the refractive index values. From the period of the optical interference of a single light source, it is in principle impossible to determine the refractive index value accurately. This is further explained in Section~\ref{sec:theory}. \citet{Berland1995} and \citet{Westley1998} solved this problem by analyzing both the period and the intensity of the reflected light. However, this method relies on accurate values of the refractive index of the substrates used (Al$_2$O$_3$ for \citet{Berland1995} and gold for \citet{Westley1998}), which are not well-characterized at astronomically relevant temperatures. \citet{Brown1996} and \citet{Dohnalek2003} determined the ice thickness using separate calibrations and obtained the refractive index values using both the period of the interference and the ice thickness. The thickness determination from separate calibrations usually possesses an uncertainty that is too large for the error allowance of refractive index values. In all the four studies mentioned \citep{Berland1995, Brown1996, Westley1998, Dohnalek2003}, after obtaining the refractive index values, the Lorentz-Lorenz relation was used to estimate the density values. In this relation, the assumption is made that the molecular polarizability of water ice is independent of the structure of the ASW. Whether Lorentz-Lorenz relation can be applied to polar ice is still under debate.
\citet{Ge2017} pointed out that this approach is suitable for gases and non-polar liquids but is inaccurate for polar liquids, such as water, while \citet{Domingo2021} suggested that this relation can be used for CH$_3$OH and NH$_3$ ices, which are both polar ices. Furthermore, molecular polarizabilities in the solid state are inferred from gas phase values and not directly measured \citep{Murphy1977, Ge2017}. Instead, it is estimated from refraction index measurements using the Lorentz-Lorenz relation. However, the Lorentz-Lorenz relation may not be suitable for ASW. More recently, \citet{Kofman2019} used a Xe arc ``white'' lamp and a UV-Vis spectrometer to measure the optical interference in the UV-Vis range and to derive the refractive index of vapor-deposited water ice within this wavelength range. The advantage of this method is that it allows measurement of the refractive index value over a large wavelength range in one-go, instead of performing measurements at each wavelength separately. However, this method suffers from the same problem as described in previous studies, e.g., by \citet{Dohnalek2003}, i.e., that the density of the ice needs to be determined separately. \citet{Kofman2019} used the density value obtained by \citet{Dohnalek2003}, assuming that the Lorentz-Lorenz relation is valid for ASW, to obtain the refractive index values in the UV-Vis range for water ice grown at different temperatures. To determine the refractive index value without relying on a separate calibration of density,  \citet{Beltran2015} used a different approach in which two light sources at different incident angles are used to measure the optical interference. The advantage of this method is that it allows measuring the refractive index value without relying on separate calibrations of the density or the Lorentz-Lorenz relation. The most recent development has been reported by \citet{Stubbing2018, Stubbing2020} who developed a new UHV-compatible UV-visible spectrometer that allows simultaneous measurement of thickness and refractive index of ices. Their setup has the capability to change the incidence angle of the light without making changes to the chamber. The employed methodology allows for the determination of both the real and imaginary refractive index of ices after its deposition, being a very promising technique for analyzing the changes in the optical constant during ice processing (e.g. UV irradiation and radiolysis of ices). However, the technique requires knowledge of the real and complex components of the refractive index of the substrate onto which the ice is grown;  these data are not always available for a broad wavelength range and for cryogenic temperatures. In this study, we build upon existing work from \citet{Kofman2019} and \citet{Beltran2015}, using both a HeNe laser and Xe arc lamp at two different incidence angles to obtain the refractive index values in the  UV-Vis range without relying on the separate calibration of ice thickness. This new approach comes with the advantages of both methods and does not rely on accurate optical parameters of the substrate.

Extinction of light is another important optical property of solids. It consists of both scattering and absorption by the solid. Single crystalline water ice is transparent between 250 and 750 nm, and has a very low value in both absorption and scattering coefficients. However, this is not the case for amorphous solid water; given its irregular and porous structure, it may strongly scatter light in the previously mentioned wavelength range. Compared to the refractive index, much less is known about the extinction or scattering of vapor-deposited water ice. Prior laboratory measurements of the light scattering by water ice focused on single crystals of water ice \citep[see e.g.,][]{Pluchino1986, Bacon2000, Ulanowski2003,Shcherbakov2006,Smith2016}, and mostly with applications to atmospheric studies. To the best of our knowledge, there is no systematic laboratory work of the extinction by vapor-deposited amorphous solid water in the UV-Vis range. In this study, we monitored the decrease in light intensity with ice thickness and thus also obtained the extinction coefficient of vapor-deposited water ice at different temperatures.

In the next sections we first briefly summarize the concept of interference measurements to determine refractive indices (Section~\ref{sec:theory}). The experiment is described in Section~\ref{sec:exp}. The results and analysis are presented in Section~\ref{sec:results}. In Section \ref{sec:astro_new} we illustrate how the data presented here can be used to interpret astronomical observations of ices in Solar System. Finally, we conclude with several findings that follow from the new approach introduced here. The astronomical deliverable of this work comprises a list with the most complete refractive indices and extinction coefficients for ASW in the UV-Vis.


\section{Theory}
\label{sec:theory}
\begin{figure}
\plotone{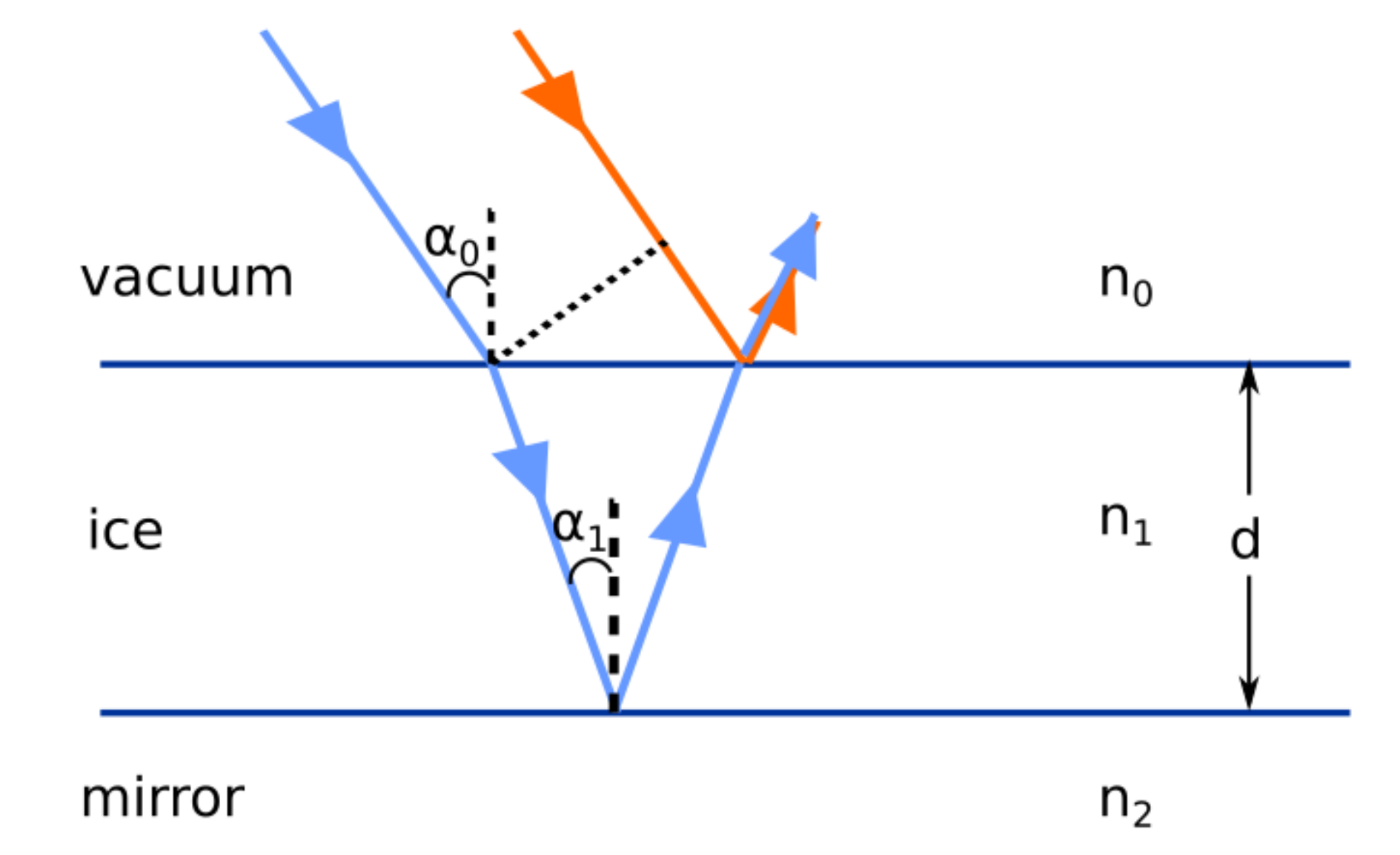}
\caption{Illustration of the optical interference pathways in a growing (increasing d) ice layer. \label{fig:interf_drawing} }
\end{figure}

Light beams reflected from the vacuum-ice interface and reflected from the ice-mirror interface interfere with each other and form periodical constructive and destructive oscillations as the ice thickness increases (see Fig.~\ref{fig:interf_drawing}).  Following a similar calculation as in \citet{Goodman1978}, we can describe the intensity of light at a certain wavelength $\lambda$ as follows:
\begin{equation}
    I=A+B\cos(\frac{2\pi \delta}{\lambda})
    \label{eq:I1}
\end{equation}
where $A$ and $B$ are two constants, $\lambda$ is the wavelength, and $\delta$ is the path difference between light beams that are reflected from the two interfaces (the blue and orange trace in Fig.~\ref{fig:interf_drawing}). It can be shown that:
\begin{equation}
    \delta = 2 n_1 d \cos(\alpha_1)
    \label{eq:delta}
\end{equation}
where $n_{1}$ is the refractive index of the ice film, $d$ is the thickness of the ice, and $\alpha_1$ is the angle between the light in the ice film and the surface normal (see Fig.~\ref{fig:interf_drawing}). Assuming that the ice is grown at a constant rate $\phi$, the thickness can be expressed as $d = \phi t$.

Combining Eq.~\ref{eq:I1} and Eq.~\ref{eq:delta}, we get:
\begin{equation}
    I=A+B\cos(\omega t)
    \label{eq:I2}
\end{equation}
where the frequency of the oscillation $\omega$ is:
\begin{equation}
    \omega = \frac{4n_{1}\pi \cos(\alpha_{1})\phi }{\lambda}
    \label{eq:omega}
\end{equation}
The value of $\alpha_{1}$ is unknown, and cannot be measured directly. However, the value of $\alpha_0$ (see Fig.~\ref{fig:interf_drawing}) is known. Using the Pythagorean identity and Snell's law, the above equation can be rewritten so that it depends on $\alpha_0$:

\begin{equation}
 \omega= \frac{4n_{1}\pi\phi }{\lambda} \cdot \sqrt{1-\left(\frac{n_{0}\sin{(\alpha_{0})}}{n_{1}}\right)^2}
 \label{eq:omega_2}
 \end{equation}

In the above equation, there are two unknown parameters $\phi$ and $n_1$. From fitting the oscillation of one light source, one cannot solve the values of both $\phi$ and $n_1$ simultaneously. With two light sources of the same frequency at different incidence angles, however, there are two equations and therefore the values of both $\phi$ and $n_1$ can be obtained. Suppose the two incidence angles are $\alpha$ and $\beta$, respectively, one gets:
\begin{align}
    \omega_{\alpha} &= \frac{4 n_1\pi \cos(\alpha_1) \phi }{\lambda}   \\
    \omega_{\beta}  &= \frac{4 n_1\pi \cos(\beta_1) \phi }{\lambda}
    \label{terms}
\end{align}
where $\lambda$ is the wavelength of the HeNe laser--- 632.8 nm. By dividing the two equations above and utilizing Eq.~\ref{eq:omega_2}, the following equation can be obtained:
\begin{equation}
    \frac{1-\left(\frac{n_0}{n_1}\right)^2 \sin{(\alpha_0)}^2}{1-\left(\frac{n_0}{n_1}\right)^2 \sin{(\beta_0)}^2} = \left(\frac{\omega_{\alpha}}{\omega_{\beta}}\right)^2
\end{equation}

Rewriting this equation gives the refractive index $n_1$ as a function of the incidence angles and the frequencies of the oscillations.

\begin{equation}
    n_1=n_0\sqrt{ \frac{\omega _\beta^{2}(\sin \alpha_0)^{2}-\omega _\alpha^{2}(\sin\beta_0)^{2}}{\omega _\beta^{2}-\omega _\alpha^{2}}}
\label{eq:n1}
\end{equation}

It is clear from this equation that the $n_1$ value only depends on the ratio between the two frequencies, but not on their absolute values. Consequently, in Eq.~\ref{eq:I2} we can also write the cosine term of the light intensity as $\cos (\omega^{'} d)$ where $\omega^{'}$ is the frequency of the oscillation when the signal is plotted as a function of thickness, and $d$ is the thickness. In practice, monolayer (ML) is used as the unit of the thickness, and the frequency is expressed in the unit of ML$^{-1}$.
From Eq.~\ref{eq:n1}, the deposition rate $\phi$ and the refractive index for $\lambda = 632.8$ nm  can be obtained simultaneously. Next, the calculation can be extended to the whole wavelength range by using the following equation derived from Eq.~\ref{eq:omega}:
\begin{equation}
    n_{2}(\lambda_{2}) = n_{1}(\lambda_{1}) \cdot \frac{\lambda_{2} \omega_{2}}{\lambda_{1} \omega_{1}}
\end{equation}
where the subscript 1 stands for the parameters at 632.8 nm and subscript 2 stands for the wavelength we want to calculate and that is easily accessible because of the broadband nature of the second (Xe arc lamp) light beam.

\section{Experimental} \label{sec:exp}
\begin{figure}
\plotone{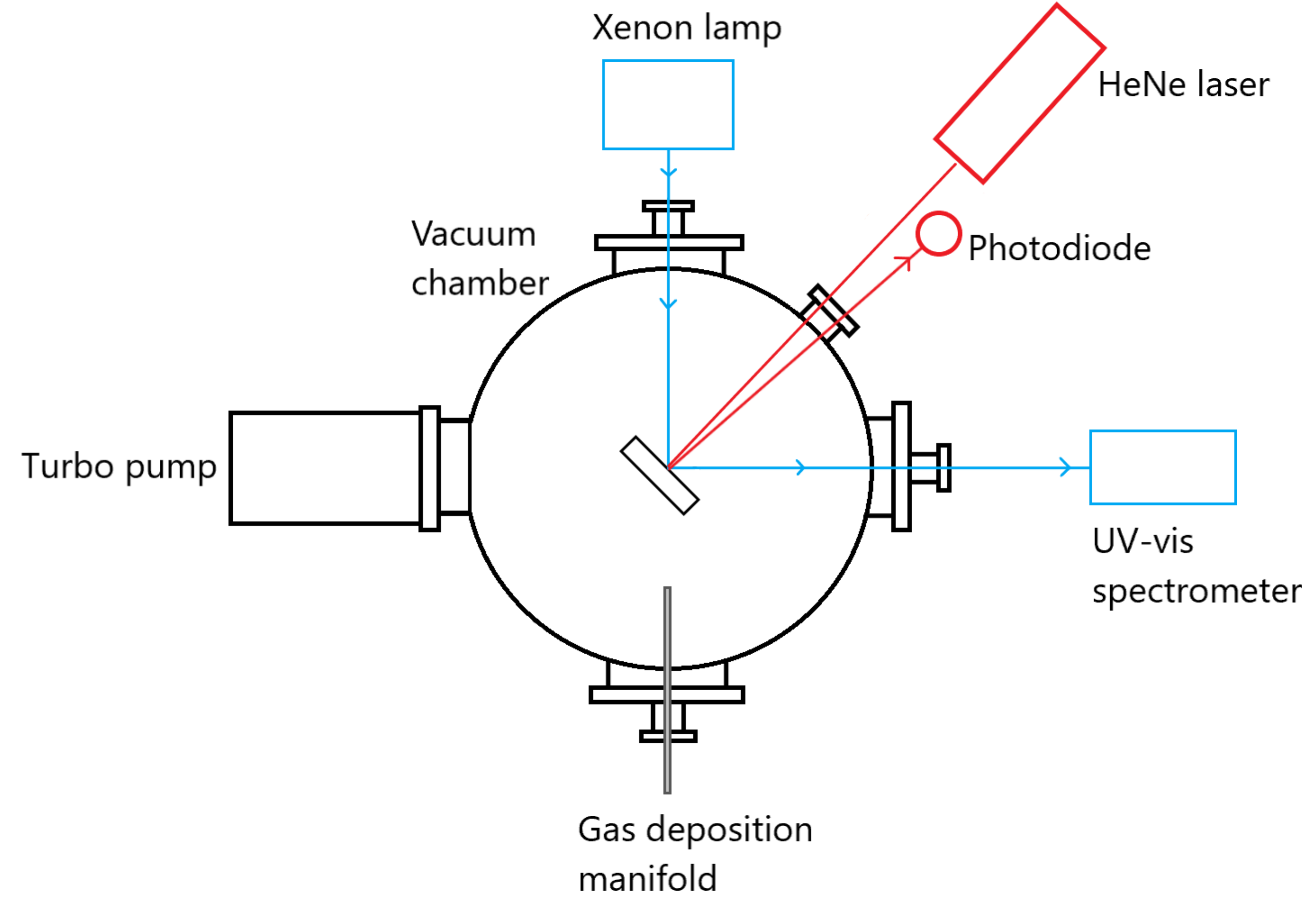}
\caption{Schematic view of the setup illustrating how the monochromatic and broadband light beams are aligned for the interference measurements. \label{fig:setup}}
\end{figure}

Experiments are performed using a high-vacuum setup located in the Laboratory for Astrophysics at Leiden Observatory. A schematic is shown in Fig.~\ref{fig:setup}. The setup has previously been described by \citet{Kofman2019}. Here we only briefly summarize the main features and the modifications to the setup. The main vacuum chamber is pumped by a turbomolecular pump to a base pressure of $7\times10^{-8}$ mbar while the cryostat is off. The pressure in the chamber is measured by a cold cathode ionization pressure gauge to an absolute accuracy of $\pm30\%$. Located at the center of the chamber is a 19 mm diameter circular UV-enhanced aluminum mirror (Thorlabs Inc.). According to the manufacturer's specifications, the flatness of the mirror surface is $\lambda/10$.
The mirror is attached to the cold finger of a closed-cycle helium cryostat that allows reaching temperatures as low as 9 K.
The temperature of the cold finger is measured by a calibrated Chromel-AuFe 0.07\% thermocouple and controlled by resistive heating. A Lakeshore 330 temperature controller ensures the absolute temperature to an accuracy of $\pm4$ K. Water ice is grown on the mirror by background deposition. The vapor pressure is manually controlled by a variable leak valve to ensure a stable pressure of $5.00\pm0.05\times10^{-5}$ mbar during the ice growth. The typical thickness of the ice is more than 20 $\mu$m. Two light sources, a HeNe laser and a Xe-arc lamp, are used for the optical interference measurement. The light emitted by the HeNe laser (632.8 nm) reflects from the mirror surface at an incidence angle of 4.4\degree in vacuum, and then its intensity is measured by a photodiode. A National Instrument data acquisition card measures the voltage of the photodiode every 300 or 500 ms. The Xe-arc lamp light reflects from the mirror surface at a 45$\pm$0.5\degree incidence angle in vacuum and then enters a UV-Vis spectrometer. Because of the relative long path of the light (both HeNe laser and UV-Vis) and small entrance of both detectors, the fraction of scattered light that enters the detectors is negligible. In the experiments, the absorption coefficients of water ice, amorphous or crystalline, are small over our wavelength range and the observed extinction is dominated by scattering. The spectrometer measures a spectrum every 1 or 2 seconds in the $\sim$250 to $\sim$750 nm range. A total of 1024 data points are evenly spread out over the full wavelength range. The UV-Vis signal is in the transmittance unit (the fraction of incident power that reaches the detector) and the ambient background light has already been subtracted.

In this study, two sets of experiments are performed. In the first set, ice is grown while the mirror remains at a constant, preset temperature. During the ice growth, the reflected light intensity shows a periodic constructive and destructive interference pattern. The frequency of the interference wave is used to derive the refractive index value of the ice as explained in Section~\ref{sec:theory}. The temperature of ice growth explored in this set is from 30 K to 160 K in 10 K steps. No experiments are performed above 160 K, because the thermal desorption rate of water ice becomes too high. 
It should be noted that during ice growth below 20 K, ASW ice tends to crack easily, changing its structure \citep{Bu2016a, Bu2016b}. For this reason no values are reported below 20 K. A number of experiments have been systematically repeated to verify reproducibility. It is found that the experiments are repeatable with only small deviations. The main source of error is from the uncertainty in measuring the incidence angle of the UV-Vis light. The estimated error for the refractive index is less than 0.01. In the second set of experiments, water ice is grown at different temperatures and subsequently heated to 160 K, to study the temperature dependence of the extinction coefficient during warm-up. We also investigated how the temperature ramp rate affects the extinction coefficient.

\section{Results analysis and discussion}
\label{sec:results}
\subsection{Refractive index}
We take the experiment in which the ice was grown at 90 K as a representative example to illustrate the data analysis procedure, and then apply the same procedure to all other deposition temperatures to calculate the corresponding refractive index values.

\begin{figure}
\plotone{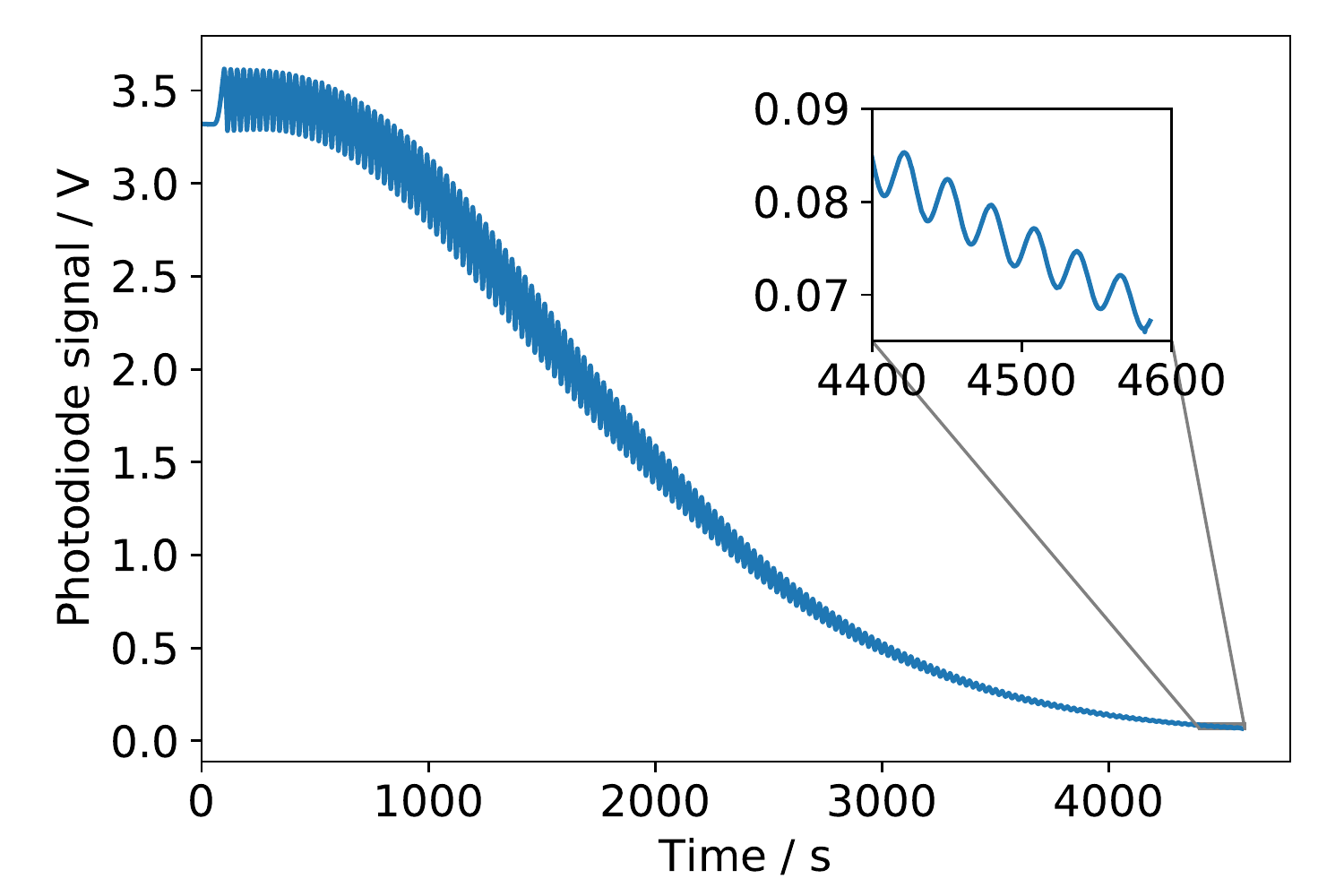}
\caption{Photodiode signal of the HeNe laser beam measured during the deposition of water vapor when the mirror was at 90 K. The inset shows the zoom-in of the region between 4400 and 4600 s.\label{fig:90k_PD} }
\end{figure}

\begin{figure}
\plotone{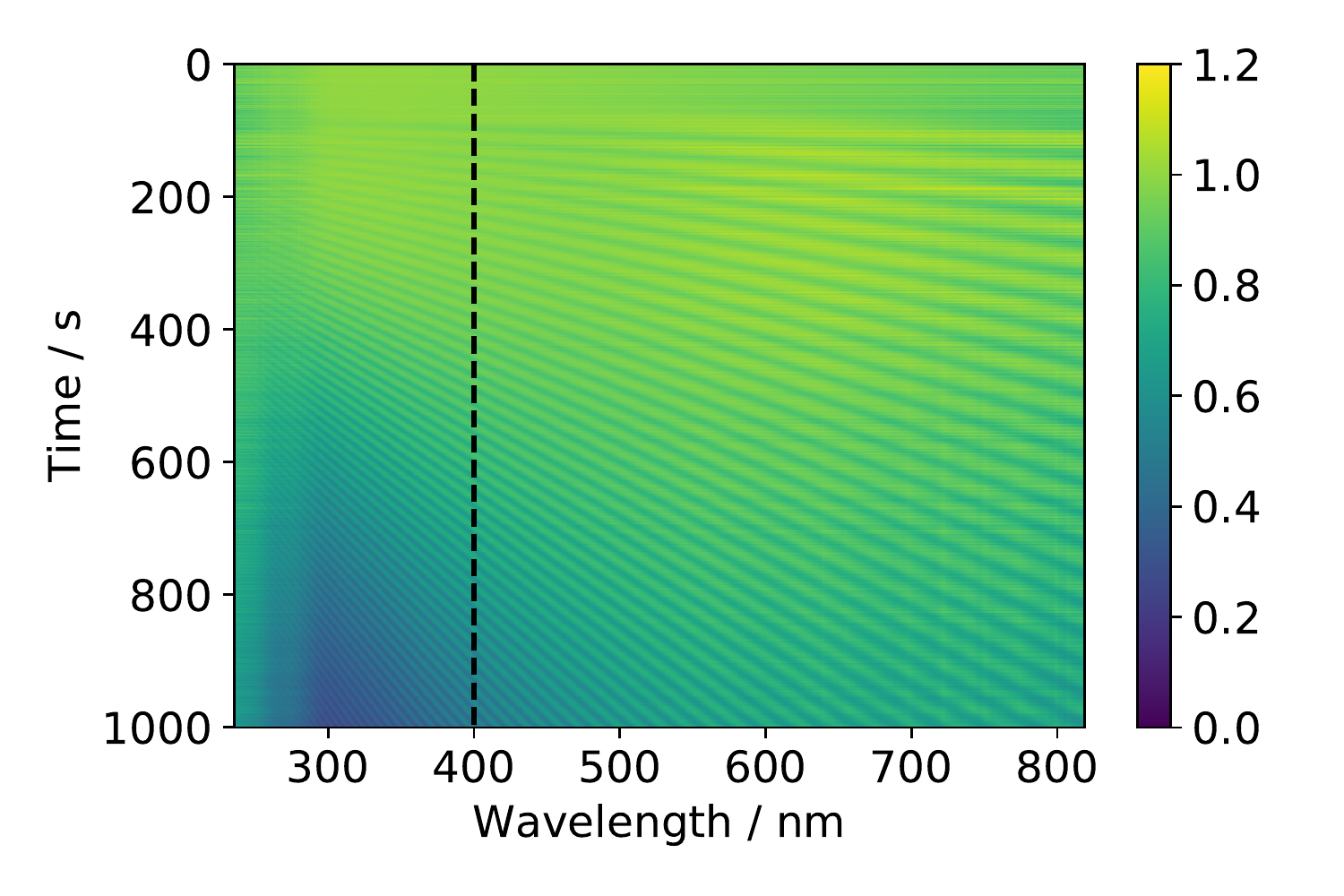}
\caption{The UV-Vis signal measured during the first 1000 seconds of water vapor deposition when the mirror was at 90 K. The data for 400 nm, which is used later on to demonstrate the performance of the method introduced here, is marked with a vertical line. \label{fig:90k_uvvis2d}}
\end{figure}

Fig.~\ref{fig:90k_PD} shows the HeNe photodiode signal during the ice growth at 90~K. A constructive and destructive interference pattern is superimposed on a baseline which decreases over time. The inset shows the zoom-in of the last 200 s of the measurement. The signal-to-noise ratio is excellent even after $\sim$150 fringes. Since the uncertainty of the oscillation frequency is inversely proportional to the number of fringes, it is desirable to record as many fringes as possible to have the most accurate refractive index values. For this reason, the refractive index measurements in this study have an accuracy roughly 10 times better than most earlier studies reported in the literature. As the deposition continues, the amplitude of the oscillation decreases. This extinction, or the decrease in overall intensity, is dominated by the scattering of light in the ice, since the absorption strength is proportional to the imaginary part of the refractive index, which is very low in the UV-Vis range \citep{Warren2008}. Here we assume that the absorption strength for amorphous water ice is similar to crystalline water ice, or at least that the absorption strength values are not are not substantially different. It will become clear later that absorption still contributes to the light extinction to a certain degree, particularly in the longer wavelength range. Fig.~\ref{fig:90k_uvvis2d} shows part of the UV-Vis broadband data for the same experiment as shown in Fig.~\ref{fig:90k_PD}. We only show UV-Vis data for the first $\sim1000$ seconds of the experiment, because it is hard to see the interference fringes visually when shown in full. Each vertical line provides an interference pattern for a specific wavelength. The frequency of the fringe pattern changes with wavelength; the longer the wavelength, the lower the number of fringes. The method of calculating the refractive index from the data has been presented in Section~\ref{sec:theory}. The next step of the analysis is to derive the frequency of the oscillations for both the HeNe photodiode signal and the UV-Vis signal at each separate wavelength. As is discussed in Section~\ref{sec:theory}, the unit of $\omega$ is s$^{-1}$ (see Eq.~\ref{eq:omega}). However, because there is a small variation in the vacuum pressure value during deposition and consequently the frequency, to smooth out the impact of pressure instability, we first plot the interference signal versus thickness in monolayer (ML, defined as 10$^{15}$ molecule cm$^{-2}$), and then determine the frequency in the unit of ML$^{-1}$. The thickness calculation is based on the impingement rate as demonstrated by \citet{He2018co2}.

\begin{figure*}
\gridline{\fig{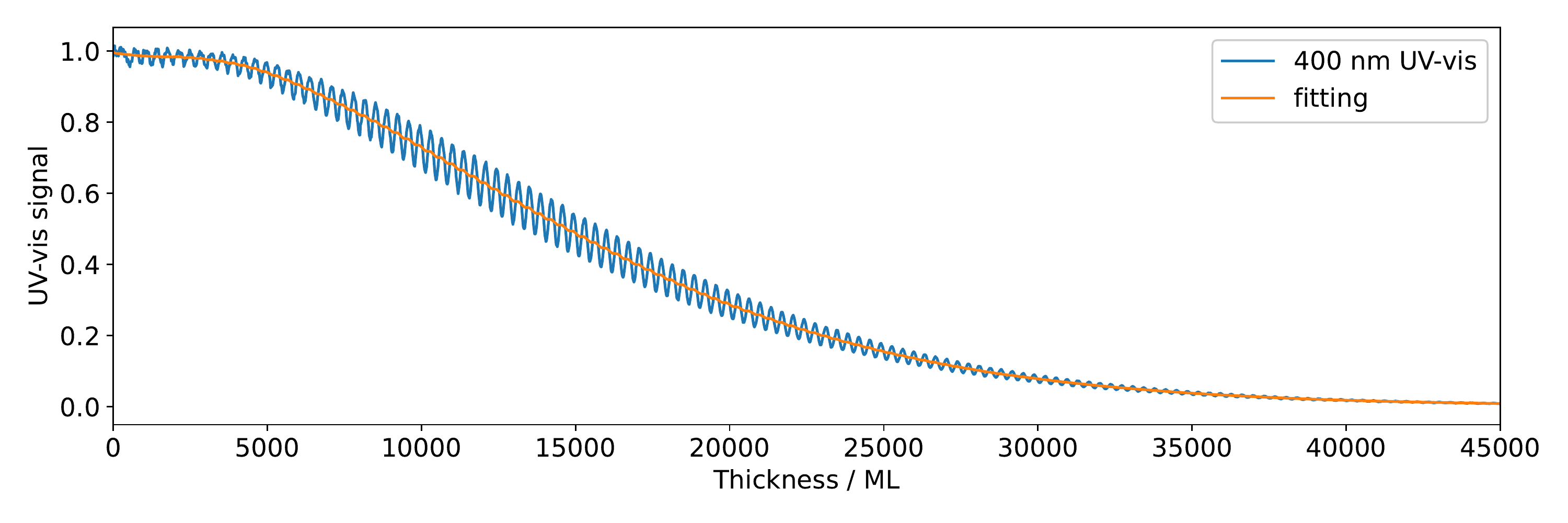}{0.9\textwidth}{(a)}
          }
\gridline{\fig{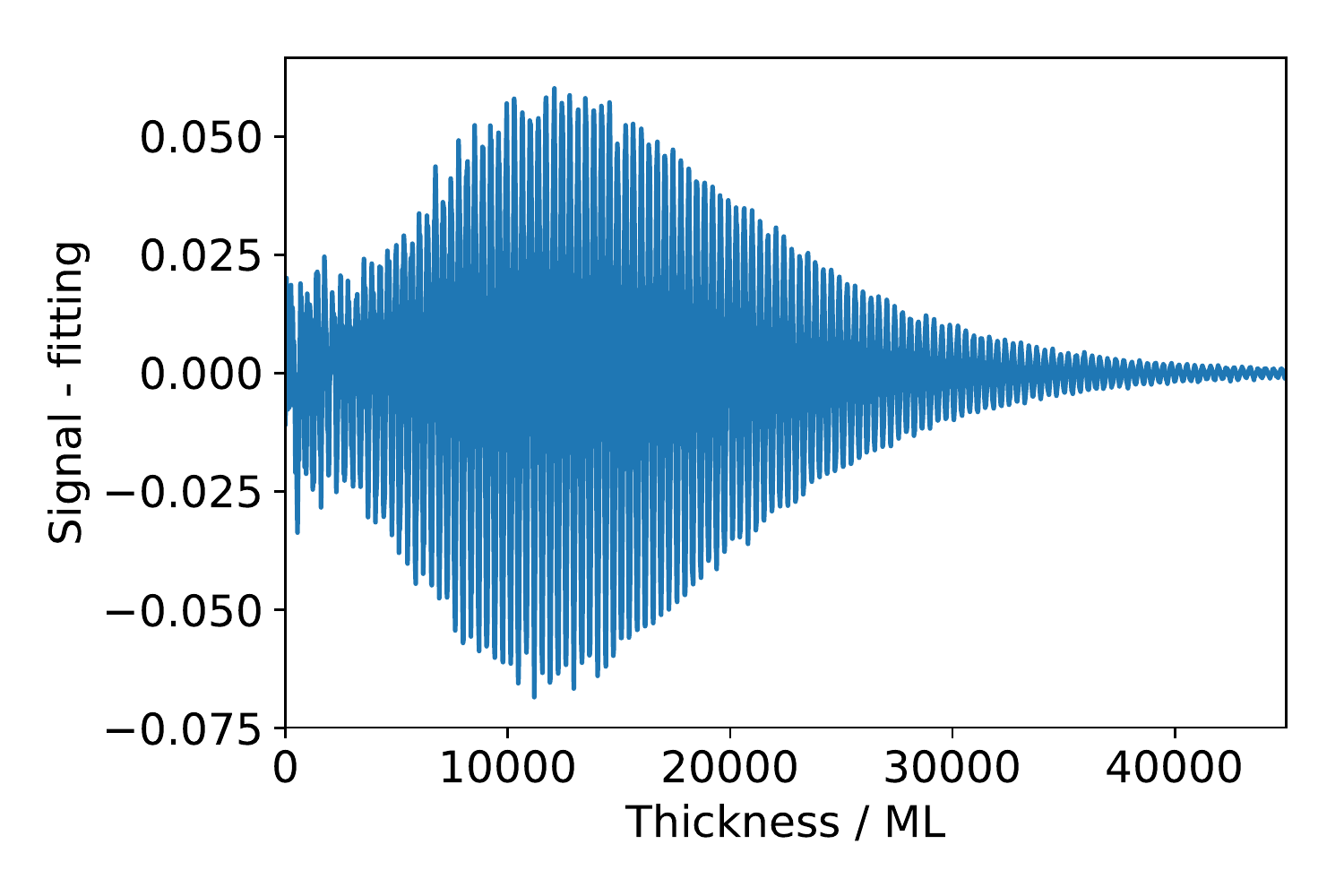}{0.45\textwidth}{(b)}
          \fig{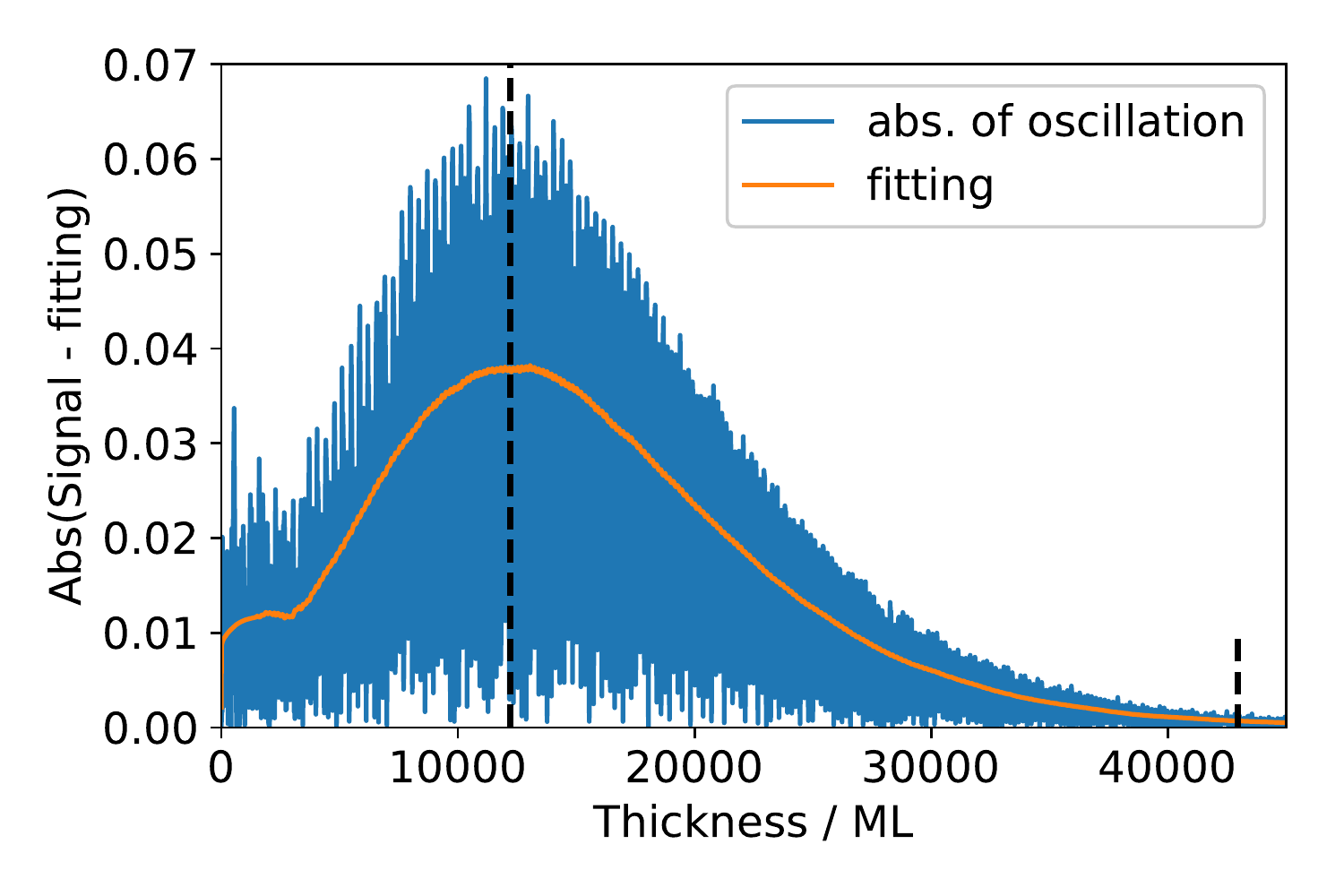}{0.45\textwidth}{(c)}
          }
\gridline{\fig{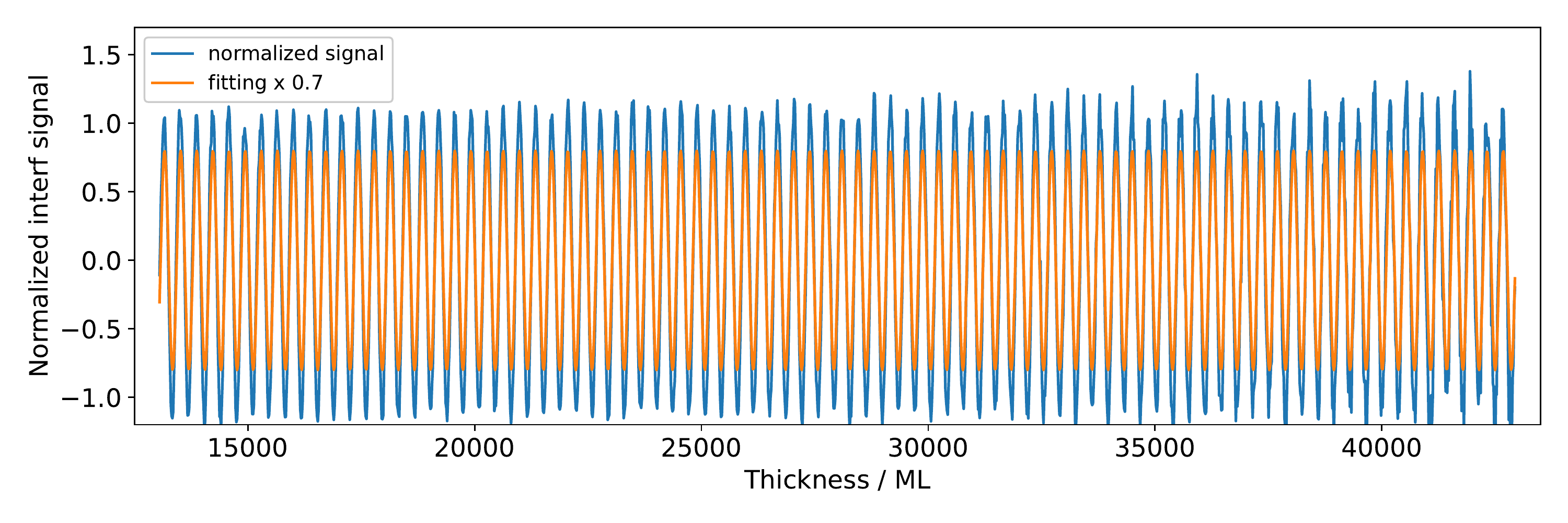}{0.9\textwidth}{(d)}}
\caption{Illustration of the data analysis procedure to obtain the oscillation frequency of the UV-Vis signal. The signal at 400 nm for the 90 K deposition experiment is used as an example. Step 1: The UV-Vis signal during ice growth is plotted against ice thickness in the unit of ML (blue curve in panel (a)); then a Savitzky–Golay filter is applied to obtain the smoothed curve (the fitting; orange curve in panel (a)). Step 2: The oscillation part of the signal in panel (a) is obtained by subtracting the smoothed curve (orange) from the original UV-Vis signal (blue); the result is shown in panel (b). Step 3: The absolute value of the curve in panel (b) is plotted, and then a Savitzky–Golay filter is applied to obtain the fitting (orange curve in panel (c)); the fitting is 1/$\sqrt{2}$ times the amplitude of the oscillation signal; the vertical line on the left marks the maximum of the amplitude and the vertical line on the right marks the position where the amplitude is 0.001. Step 4: The section of the data between the two vertical lines is extracted, and then divided by the amplitude of the oscillation to obtain the normalized oscillation, as shown in panel (d); the normalized oscillation is then fitted with a cosine function to obtain the frequency value; for clarity purpose, the fitting is multiplied by 0.7.
\label{fig:uvvis_process}}
\end{figure*}

In the UV-Vis data, there are in total 1024 wavelength values spread over the whole wavelength range. The frequency of each fringe pattern needs to be calculated. We take the interference pattern at 400 nm as an example (the corresponding data is indicated in Fig.~\ref{fig:90k_uvvis2d} with a dashed vertical line) to illustrate the data analysis procedure, as shown in Fig.~\ref{fig:uvvis_process}. Panel (a) shows the UV-Vis signal at 400 nm along with the smoothed curve (the fitting) obtained by applying a Savitzky–Golay filter. The filter removes the oscillation and keeps the trend of the baseline. The oscillation part of the signal is obtained by subtracting the fitting from the original signal and is shown in panel (b). We analyze the oscillation frequency versus thickness and find that at the beginning of the ice growth, the frequency is unstable. This is also seen from the lower signal-to-noise ratio in the first few fringes in panel (a). The first few fringes are generally unsuitable for refractive index calculations. In similar laboratory studies in the literature, typically only a few fringes are measured, and therefore there might be a relatively high uncertainty in the derived refractive index values. Toward the end of our ice deposition, the UV-Vis signal becomes too weak, and the uncertainty of the frequency calculation becomes large. The middle part of the data is more reliable and sufficiently large for frequency determination. We select the point where the amplitude is maximum as the starting cutoff, and the point where the amplitude drops to 0.001 as the ending cutoff. Panel (c) shows the absolute value of the curve in panel (b) as well as the curve obtained by applying a Savitzky–Golay filter (labeled as ``fitting'' in the figure). The amplitude of the oscillation is about $\sqrt{2}$ times the fitting curve. The figure also shows the starting and ending cutoffs in vertical dashed lines. The next step is that we extract the data between the two cutoff points and divide them by the amplitude of the oscillation to obtain the normalized oscillation, which is shown in panel (d). Afterward, a cosine function is used to fit the normalized oscillation to obtain the frequency of the oscillations. The fitting is shown in panel (d) in orange color. In the figure, we multiplied the fitting by 0.7 to distinguish from the normalized oscillation signal.

\begin{figure*}
\plotone{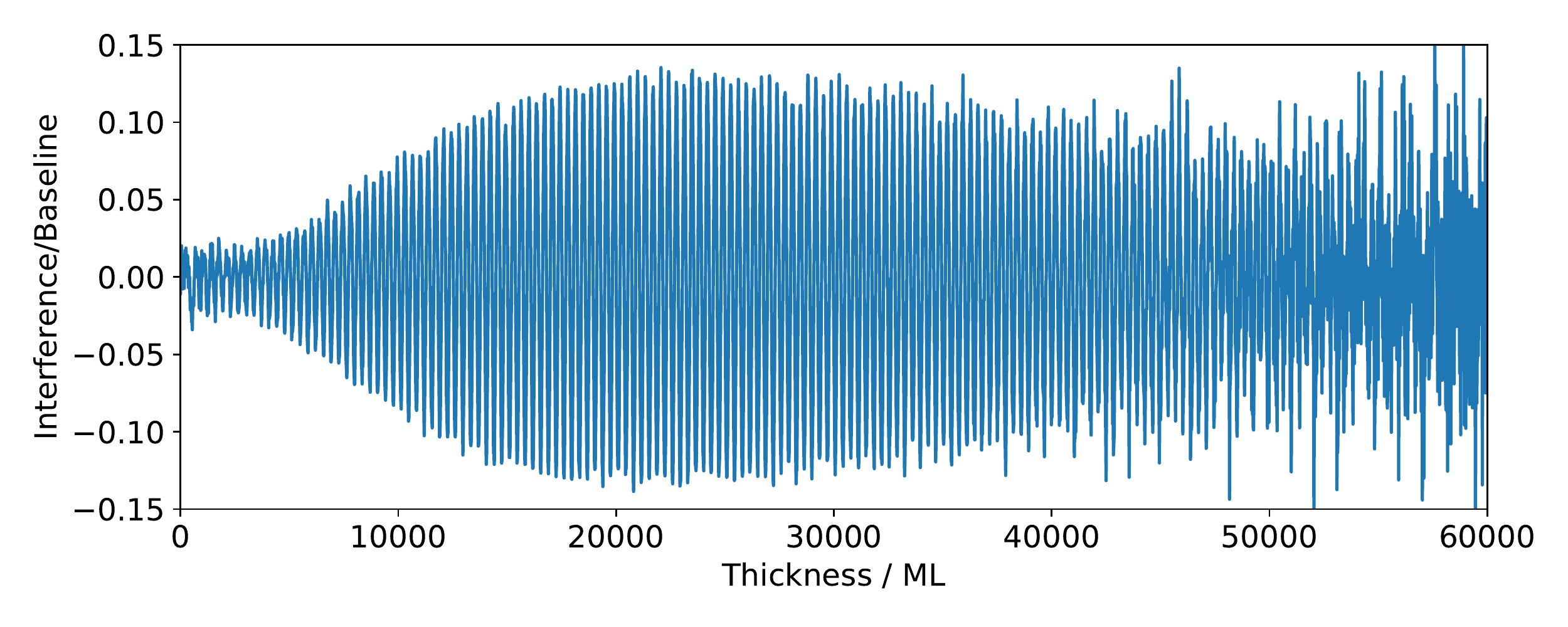}
\caption{The oscillation part of the signal as shown in Fig.~\ref{fig:uvvis_process}(b) divided by the fitting curve in Fig.~\ref{fig:uvvis_process}(a). \label{fig:90k400nm_Interfere_div_Baseline}}
\end{figure*}

In Fig.~\ref{fig:uvvis_process} panel (b), the amplitude of the oscillation increases at the beginning and then drops after about 12000 ML. Ideally, the amplitude should be proportional to the overall intensity if the ice is uniform. We divide the oscillation part of the signal (panel (b)) by the trend (orange curve in panel (a)), as is shown in Fig.~\ref{fig:90k400nm_Interfere_div_Baseline}. In the first 20000 ML, the amplitude increases, and then it decreases slowly afterward. The spikes near the end of the experiment are due to the low signal-to-noise ratio. The increase at the beginning is likely due to the impact of the surface, i.e., non-uniformness of the ice thickness at the beginning of the ice growth. However, based on the specifications of the mirror, the thickness variation is in the order of 10\% of the wavelength, i.e., a few tens of nanometers, or a few hundred ML at most. Furthermore, a Monte Carlo simulation of vapor-deposited ASW by \citet{He2019asw} has shown that if ice is grown on a flat surface, the structure is more or less uniform after 10 ML, which is much smaller than the 20000 ML value above. The explanation for this unexpectedly large difference is that even if the mirror roughness is only affecting the very beginning of the ice growth, during further deposition, light beams can still reach the non-uniform bottom layers. The effect of these bottom layers on the full pattern is only averaged out slowly with 20000 ML of ice. 
After $\sim$20000~ML, the amplitude of the normalized oscillation decreases. This is likely because more scattered light enters the UV-Vis detector and raises the ``trend'', regardless of the temperature. This observation, therefore, is an intrinsic consequence of the applied methodology. It should be noted that this only applies to the light beams. In some prior astrochemical laboratory studies \citep{Hidaka2008, He2017}, it was shown that the surface (usually a metal surface) does not (strongly) affect chemical processes on thin films of water ice, typically of the order of 100 ML, whereas the substrate performs a critical role in supporting the interference of light transmitted by the ice film.

\begin{figure*}
\plotone{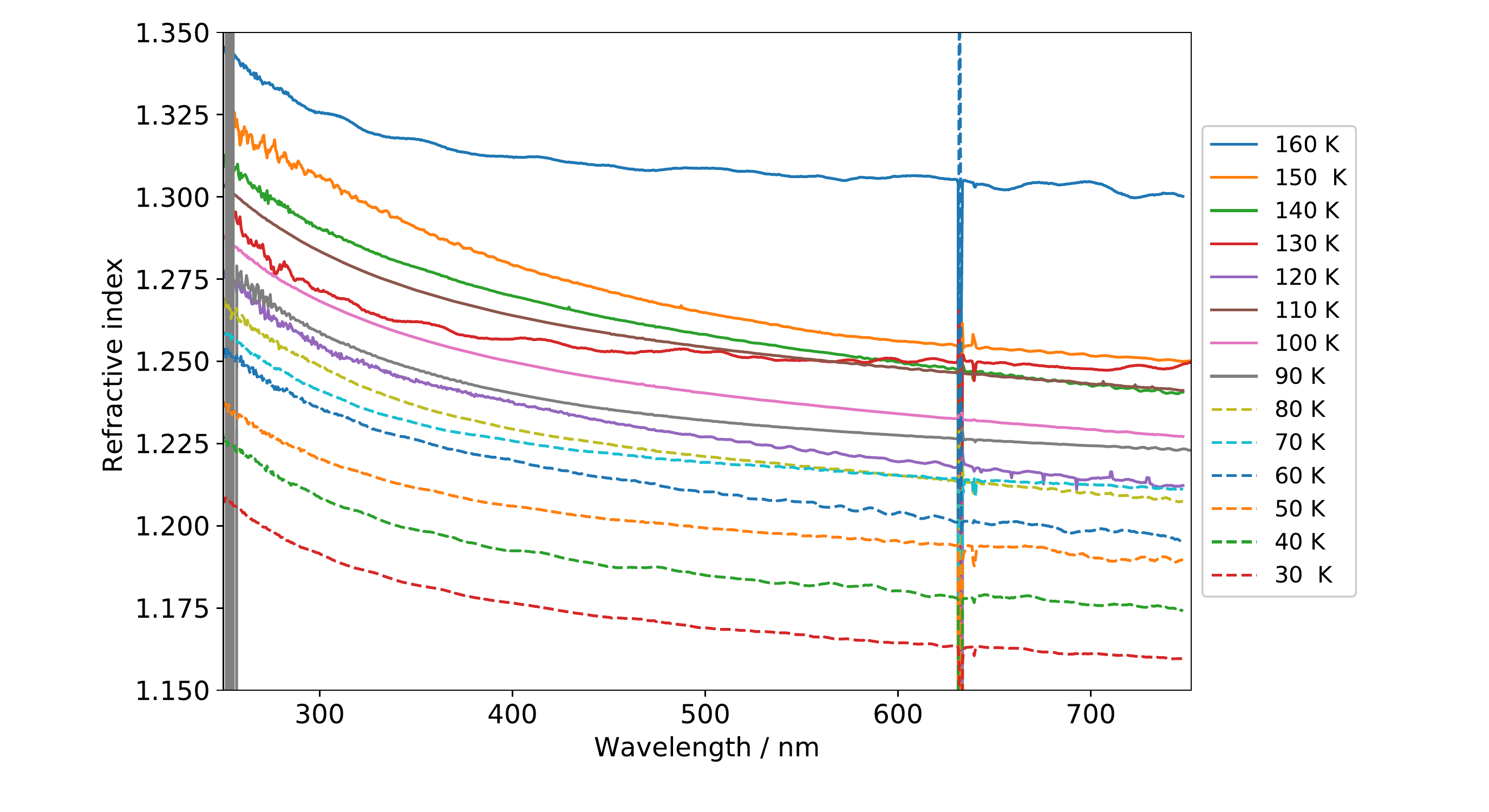}
\caption{Wavelength-dependent refractive indexes in the UV-Vis range for ice that is grown at different temperatures. Data in this are available from Leiden Ice Database (\url{https://icedb.strw.leidenuniv.nl/refrac_index}). \label{fig:allT_n_l} }
\end{figure*}

After obtaining these frequencies, we follow the calculations in Section~\ref{sec:theory} and calculate the refractive index values for wavelengths between 250 and 750 nm. It should be noted that the UV-Vis signal for wavelengths of 632.8$\pm3$ nm is affected by the HeNe laser, and therefore the oscillation frequency value in that range is calculated by averaging the data points below and above the HeNe wavelength. Similarly, the above procedure is applied to all wavelengths of the UV-Vis signal, as well as the HeNe photodiode signal. In Fig.~\ref{fig:allT_n_l}, we show the refractive index value in the UV-Vis range for water ice deposited at temperatures in the range 30--160 K. These refractive index curves generally show similar behavior, with increasing values towards the UV range. The figure also shows that the refractive values generally increase with deposition temperature. This is related to the fact that the porosity decreases with temperature. However, an exception exists at $\sim120~K$, which has a lower refractive index value than found for 110 and 130~K. This finding will be discussed later. 

The curves derived here differ from that of Fig. 6 in \citet{Kofman2019} in two ways: (1) The refractive index versus wavelength curves in Kofman et al. were calculated using the Sellmeier equation and Lorentz-Lorenz equation, assuming the same shape/slope for all deposition temperatures, while in Fig.~\ref{fig:allT_n_l}, each data point is calculated from the oscillation frequency of the interference fringes. As a consequence, the slopes for the investigated deposition temperatures are not fully comparable with those found in \citet{Kofman2019}; for example, the curves at 130 and 160 K are flatter than those derived for other temperatures. (2) The height of each curve in Fig. 6 of \citet{Kofman2019} was determined by applying a linear fitting to the refractive index values at 632.8 nm in \citet{Dohnalek2003}, while the results here do not rely on data derived from other experiments. This also explains the up to 5\% difference in absolute refractive values of the present work compared to those reported in Kofman et al.. Moreover, in Kofman et al., the refractive index values increase exactly linearly with temperature, while in this study, the linear relation does not hold, at least not for temperatures around 120 or 150--160 K.

\begin{figure}
\plotone{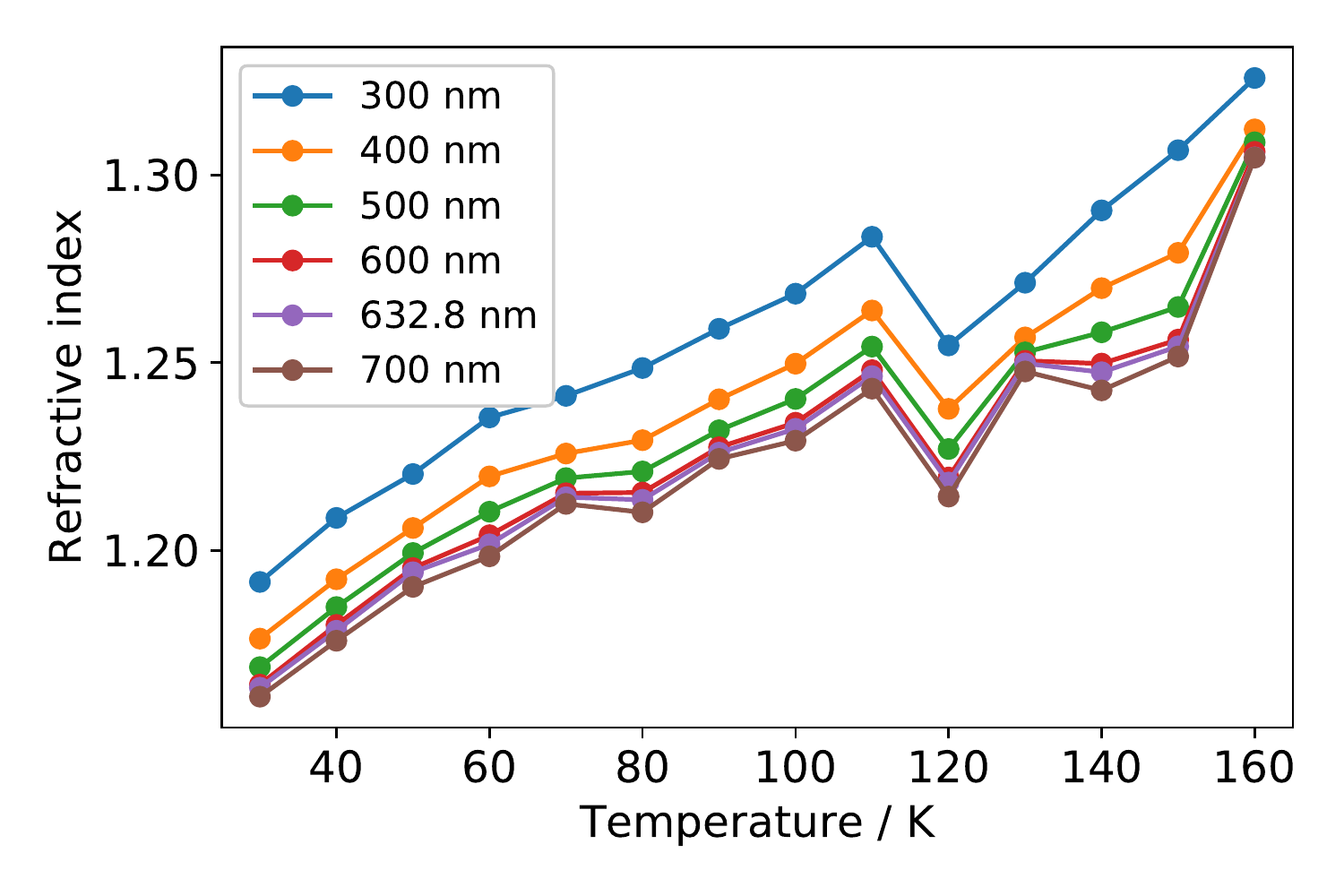}
\caption{Refractive indexes for selected wavelengths for ices grown at different temperatures.  \label{fig:n_l_selected}  }
\end{figure}

Fig.~\ref{fig:n_l_selected} shows the refractive index versus ice deposition temperature at 632.8 nm as well as a few other wavelengths. The $n_{632.8}$ value differs from that by \citet{Dohnalek2003} (on which Kofman et al. was based) in that the data point at 120 K is clearly lower than that at 110 or 130 K, and there is a significant ``jump'' between 150 and 160 K. The refractive index values at other wavelengths also show similar behavior. Although the value at 120 K is only 0.03 or 2.5\% lower than at 110 K or 130 K, it was verified to be fully reproducible in independent measurements. In Section~\ref{sec:extinction}, it is shown that this temperature (120 K) coincides with the temperature where also a large increase in the extinction coefficient occurs, which further suggests that the lower refractive index value at 120 K has a physical reason. The larger increase in the refractive index between 150 and 160 K is very likely related to the (starting) crystallization of water ice.

\subsection{Extinction coefficient}
\label{sec:extinction}
\begin{figure}
\plotone{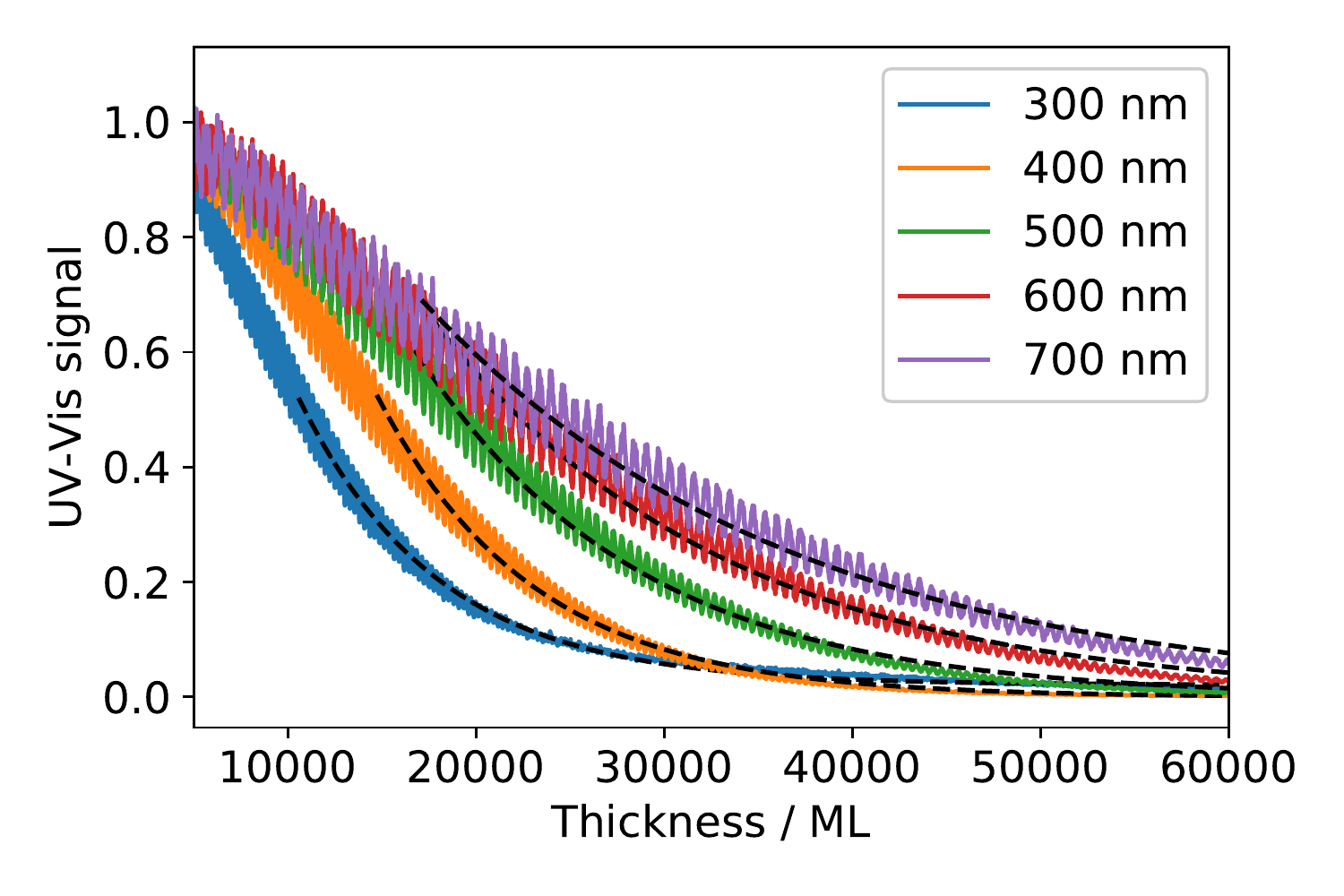}
\caption{Fitting of the UV-Vis signal for selected wavelengths using exponential decay functions in Eq.~\ref{eq:exp_decay}. The beginning part of the curve are is not taken into account in the fitting procedure (see text). \label{fig:90k_trend_fit}}
\end{figure}

As the ice becomes thicker, the measured light intensity decreases due to extinction by the ice. If the ice is uniform, the light that transmits through it should follow an exponential decay with thickness. We exclude the beginning part of the data, as shown in Fig.~\ref{fig:uvvis_process} panel (c), the part before the amplitude of the oscillation reaches the maximum, and choose the later part to analyze the exponential decay. The data are fitted with the formula:
\begin{equation}
  Intensity = A\exp(-B d) + C
  \label{eq:exp_decay}
\end{equation}
where $d$ is the thickness of the ice in ML, and $A$, $B$, $C$ are fitting parameters. Fig.~\ref{fig:90k_trend_fit} shows an example fitting of the UV-Vis signal. Given the good exponential fit, we conclude that there is significant extinction at all wavelengths. Several previous experimental studies \citep{Berland1995, Westley1998, Dohnalek2003} reported that vapor-deposited water ice is transparent and there is negligible scattering. Those studies were all limited to very thin ices that only have a few interference fringes at 632.8 nm, and therefore it is not surprising that clear scattering effects could not be observed. In our experiments, to determine reliable extinction coefficient values, ice was grown continuously until most of the light is scattered; the exact thickness value varies with the deposition temperature.

\begin{figure*}
\plotone{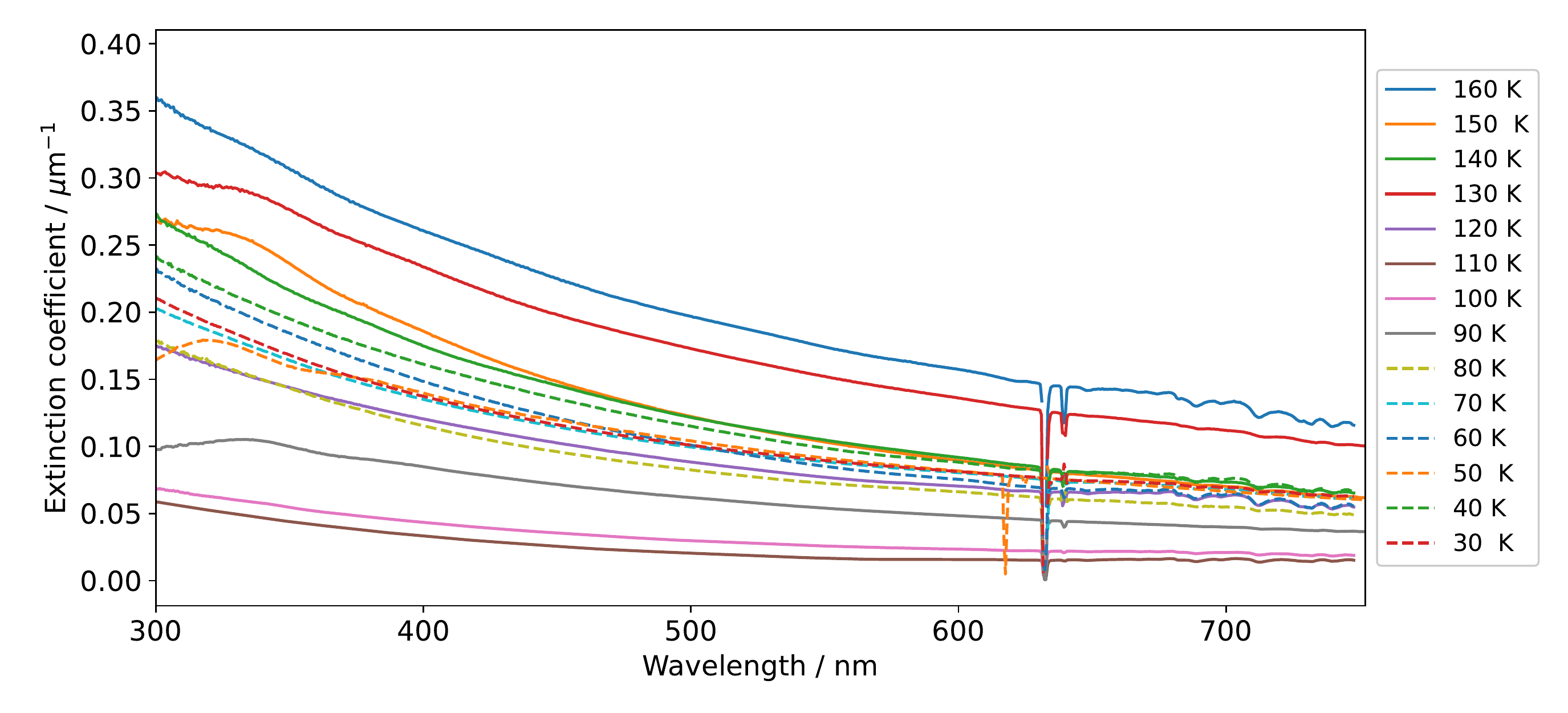}
\caption{Wavelength-dependent extinction coefficient for ice grown at different temperatures by vapor deposition. The negative spikes close to 632.8 nm are artifacts introduced by the HeNe laser. Data in this are available from Leiden Ice Database (\url{https://icedb.strw.leidenuniv.nl/refrac_index}). \label{fig:allT_decay}}
\end{figure*}

The real pathlength $d'$ that the light passes through is not the same as thickness $d$. It depends on the incidence angle in the vacuum (angle $\alpha_0$) as well as the refractive index value at that specific wavelength.  We calculate the pathlength $d'$ using the following relation:
\begin{equation}
    \frac{d'}{d} = \frac{2}{\cos (\alpha_1)}  = \frac{2}{\sqrt{1-(\frac{\sin (\alpha_0)}{n_1(\lambda)})^2}} \label{eq:dprime}
\end{equation}
When we plug this formula into Eq.~\ref{eq:exp_decay} to correct for the thickness and apply the calculation to the UV-Vis signal for each wavelength and each temperature, we obtain the extinction coefficient of vapor-deposited water ice as a function of wavelength and temperature. A conversion is also made from ML to $\mu$m based on the refractive index values in Fig.~\ref{fig:allT_n_l}. The extinction coefficient results are shown in Fig.~\ref{fig:allT_decay}. The relative error of the coefficient is estimated to be 15\% based on repeated experiments. For all deposition temperatures, the extinction coefficient decreases with wavelength. This is not surprising since it is known that light with a shorter wavelength is scattered more effectively than longer wavelength light. In the longer wavelength range, e.g., $\lambda > 680$ nm, there are some small features, which are probably due to the absorption of water ice. It is well-known that water has overtone bands of the O-H stretch and bend extending in the near infrared and beyond.

\begin{figure}
\plotone{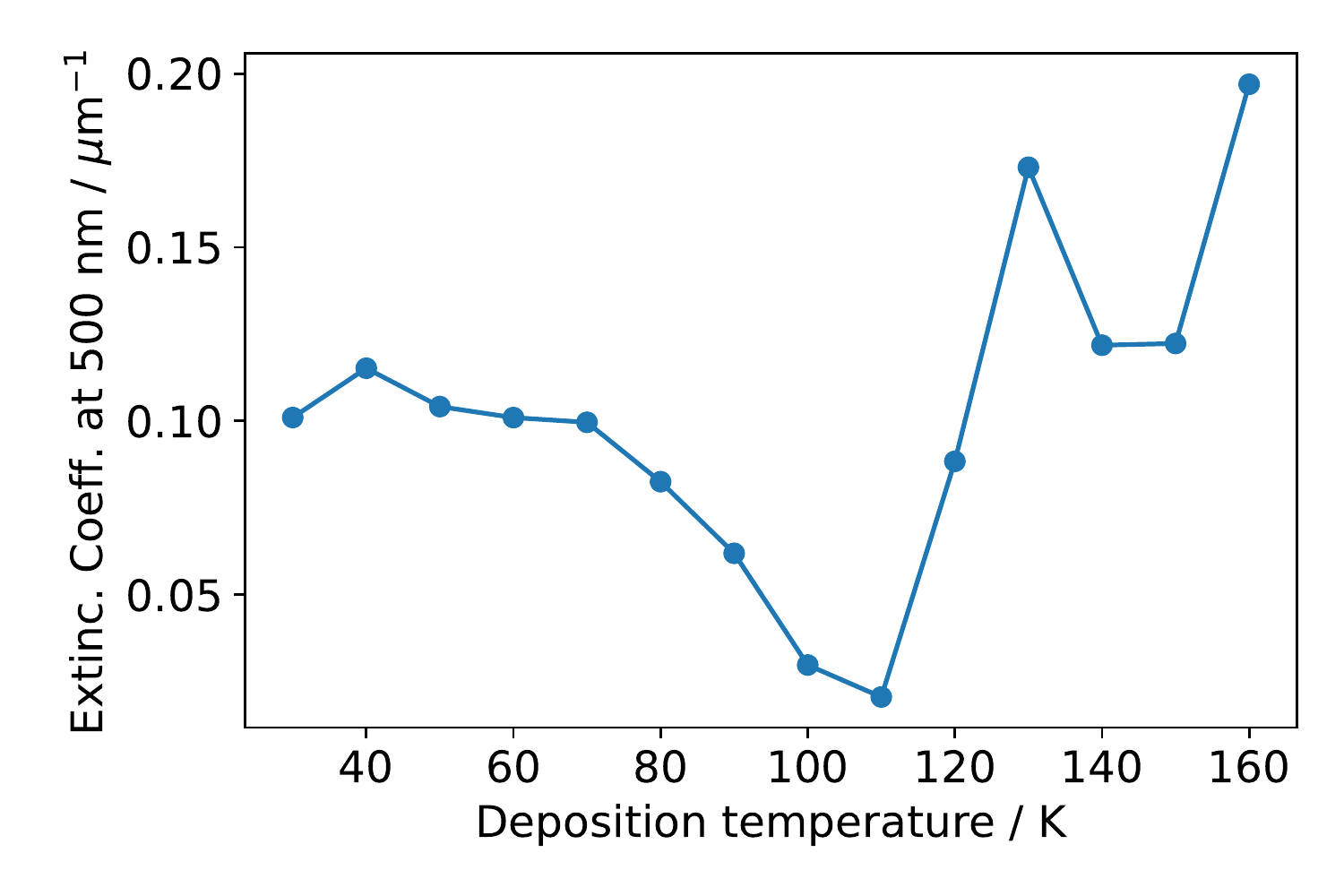}
\caption{The extinction coefficient at 500 nm for ice grown at different temperatures by vapor deposition. }
\label{fig:decay_500nm}
\end{figure}

To better see the temperature dependence of the extinction coefficient, in Fig.~\ref{fig:decay_500nm} we show the extinction coefficient at 500 nm for ice grown at different temperatures. From 30 to 110 K, the extinction coefficient decreases. By 110 K, it reaches the lowest value, i.e., here the ice has the highest transparency.

Between 110 K and 130 K, the extinction coefficient increases by a few folds. If we consider 500 nm as an example, the extinction coefficient is increased by a factor of 8.5 within only 20 K. At 130 K, the ice becomes highly opaque. The large increase of the extinction coefficient between 110 and 130 K is consistent with similar deviating behavior in the refractive index (see Fig.~\ref{fig:allT_n_l} and \ref{fig:n_l_selected}) and hints at a structural change in the ice structure. To the best of our knowledge, this behavior in the ice properties of water deposited between 100 K and 130 K was not reported before. It is not directly clear why this switch does not happen at the temperature where the crystallization of vapor-deposited water ice happens --- 155 K \citep{May2012, May2013}. Our tentative explanation is that as the ice becomes more compact, chunks of compact (amorphous) ice or polycrystalline grains of the size comparable with the wavelength of the light is forming. We reason that whenever the size of the grains, no matter if they are compact amorphous ice or small crystals, is approaching the scale of the wavelength, a large increase may be seen in the light scattering and extinction. From 130 K to 160 K, the extinction coefficient decreases and then increases, suggesting further structural changes that must be related to ice crystallization. 

The Lorentz-Lorenz relation has been used by several groups to calculate the refractive index value for ASW, either directly or indirectly \citep{Berland1995, Brown1996, Westley1998, Dohnalek2003, Kofman2019, Stubbing2020}. As pointed by \citet{Dohnalek2003}, the validity of the Lorentz-Lorenz relation is based on the assumption that the film is homogeneous on the length scale set by the wavelength of the light. This assumption requires that the light scattering due to film inhomogeneities is negligible. In this study, the high extinction coefficients measured at 130 K and also at very low temperatures suggest that the Lorentz-Lorenz relation does not always apply to ASW.

\subsection{Extinction by water ice during warming-up}
\label{sec:TPD}
\begin{figure}
\plotone{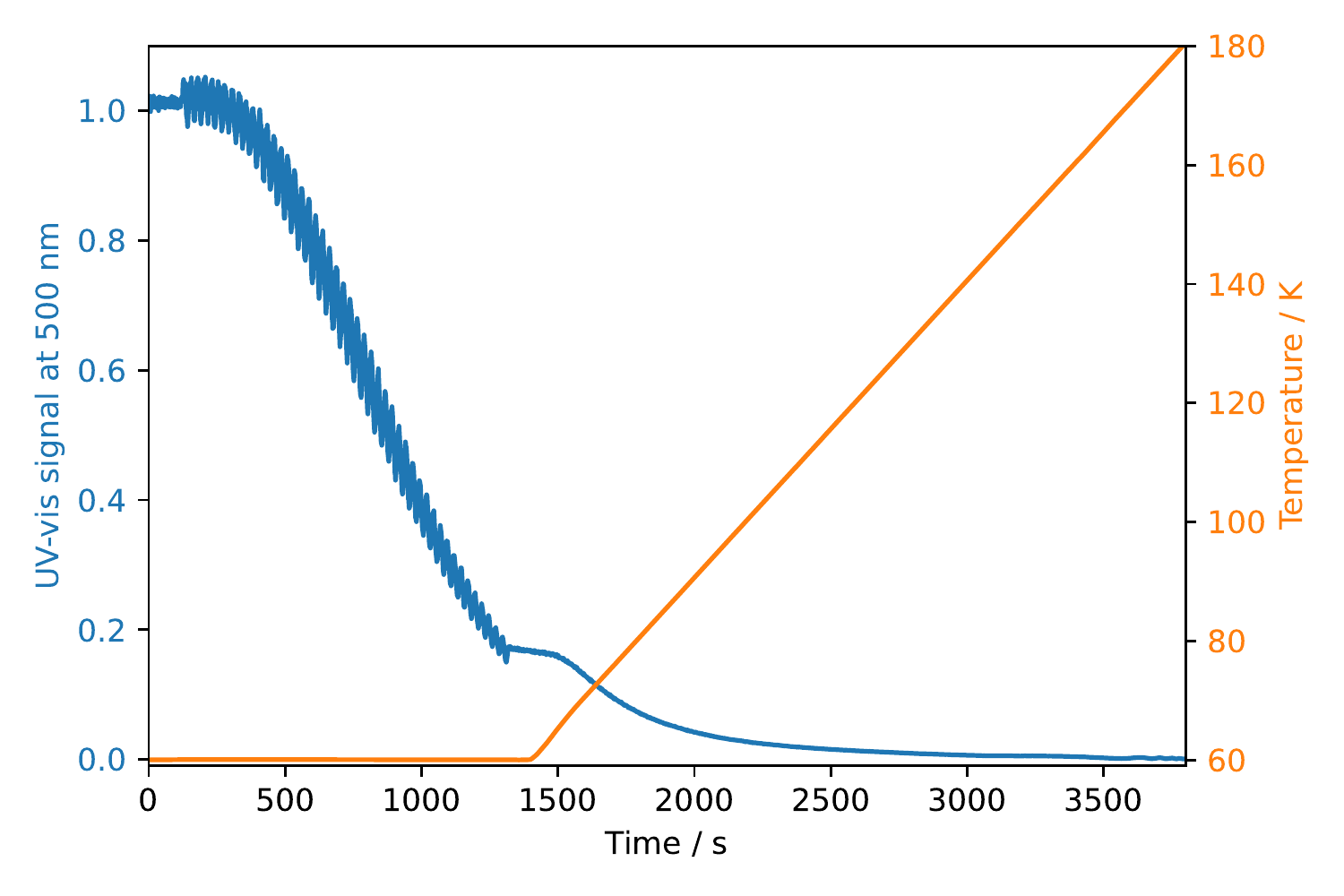}
\caption{The UV-Vis signal at 500 nm recorded in the experiment in which ice was grown at 60 K and then warmed up. The orange curve shows the temperature curve.\label{fig:TPD_60k} }
\end{figure}

\begin{figure}
\plotone{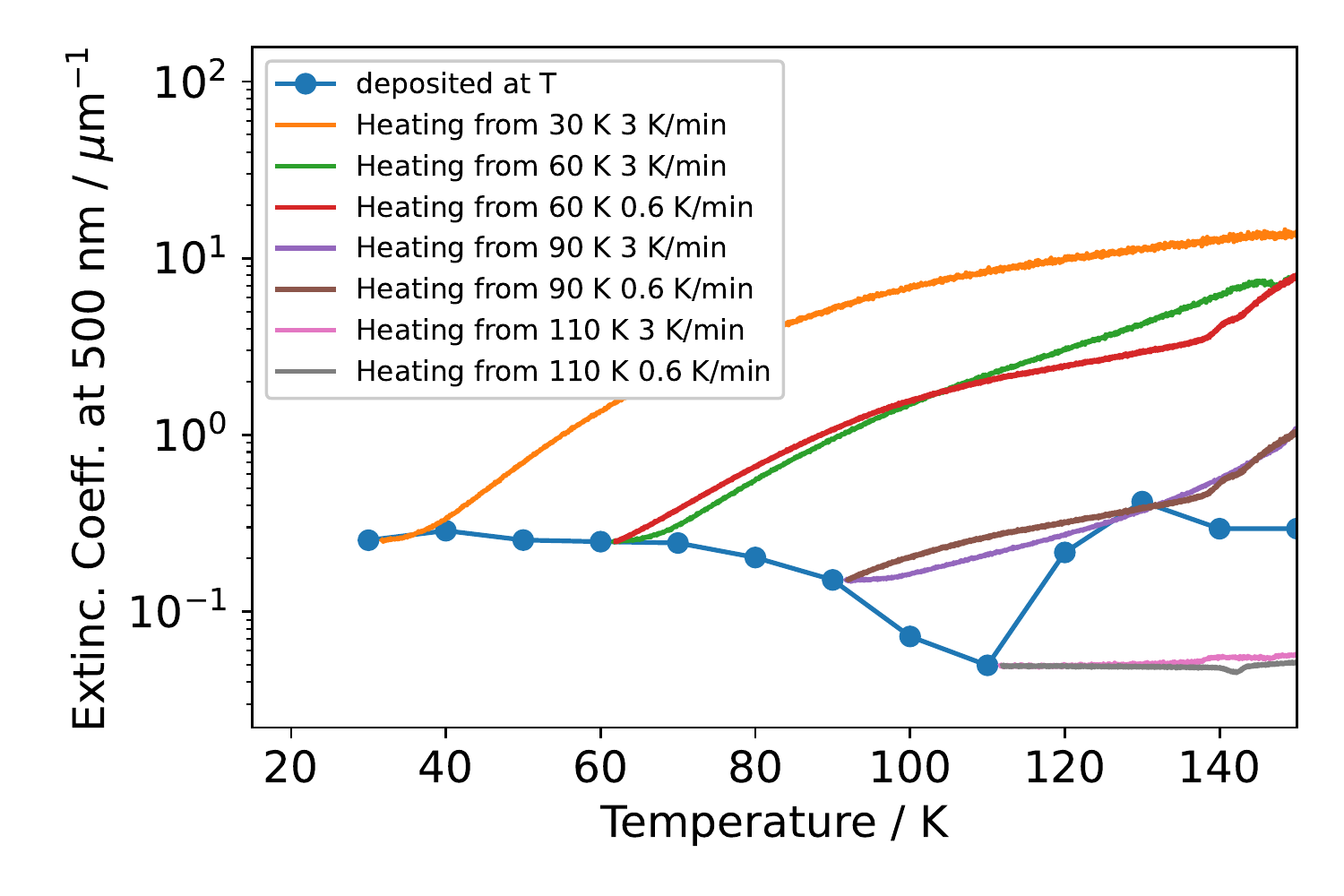}
\caption{Extinction coefficient at 500 nm for ice grown at different temperatures (blue circles), and for ice grown at 30, 60, 90, 110 K and warmed up linearly. Two different temperature ramp rates, 3 K/minute and 0.6 K/minute are used to check the impact of the ramp rate.\label{fig:TPD_compare} }
 
\end{figure}

To find out whether the temperature dependence of the extinction coefficient also applies to ice that is deposited at low temperature and warmed up, we carried out another two experiments. In the first experiment, we deposited water vapor at a pressure of $5\times10^{-5}$ mbar for 20 minutes when the substrate was set at 30, 60, 90, and 110 K, and then warmed up the ice at a ramp rate of 3 K/minute to above 150 K. During both the ice growth stage and the warming-up stage, both HeNe laser photodiode and UV-Vis signal were recorded. Between 30 K and 150 K, the effect of thermal desorption is negligible, and the ice thickness in terms of ML remains unchanged. In Fig.~\ref{fig:TPD_60k}, we show the temperature and the UV-Vis signal for 500 nm for the experiment in which the ice was grown at 60 K. In the warming-up stage, the signal at 500 nm decreases monotonically, and there is no tendency to increase in the 110--130 K range. This suggests that the significant structural difference between 110 K and 130 K is only for the ice that is deposited at these temperatures, but not for ice that is warmed up to this temperature range. We also repeated some of these experiments at a lower temperature ramp rate 0.6 K/minute to check the ramp rate dependence of the refractive index. In Fig.~\ref{fig:TPD_compare} we compare the extinction coefficients in these warming-up experiments with that from Fig.~\ref{fig:decay_500nm}. The temperature-dependent extinction coefficient during warming up is calculated by fixing the extinction coefficient at the beginning of the warming up to the one shown in Fig.~\ref{fig:decay_500nm}, and then taking the inverse of the UV-Vis signal at 500 nm. Here we ignore the change in the refractive index during warming up, and consequently, the change in pathlength $d'$ as in Eq.~\ref{eq:dprime} because the change in refractive index value only has a small impact on the extinction coefficient values shown in Fig.~\ref{fig:TPD_compare}. By comparing the curves of 3 K/minute with those of 0.6 K/minute, we find that the difference in ramp rate only has a small impact on the extinction coefficient during warm-up. From this figure, it is also clear that the structure of the ice grown at a higher temperature must be substantially different from the ice that is grown at a lower temperature and then warmed to the same temperature. For example, the extinction coefficient of the ice grown at 30 K and then warmed up to 110 K is about three orders of magnitude higher than the ice grown at 110 K, as shown in Fig.~\ref{fig:TPD_compare}. The extinction coefficient of deposited ice and the warmed ice even show an opposite temperature dependence. For instance, at a deposition temperature between 30 K and 110 K, the extinction coefficient decreases monotonically, while the ice deposited at 30 K and warmed up to 110 K shows the opposite. The ice deposited at 110 K, however, seems to remain the high transparency when warming up to higher temperatures. The reason for the $\sim$3 orders of magnitude increase in extinction coefficient when the ASW deposited at 30 K is warmed up to $\sim150$~K is still puzzling. It should also be mentioned that such an increase of extinction coefficient is also seen for the other wavelengths studied in the present experiments. The increase of extinction coefficient seems to agree with the scenario that the ASW is cracking gradually during warming up. The increase of extinction can be interpreted by the formation of icy structures within the ice with a scale comparable to or larger than the wavelength. Our experiments then suggest that the warming up of the ice allows for some cracking mechanisms similar to the one observed in \citet{Bu2016a} which would gradually form large icy structures. The experiments by \citet{Bu2016a} showed that ASW spontaneously cracks during growth if a certain critical thickness is reached. The higher the temperature, the larger the critical thickness. We have observed similar cracking properties in our experiments during the growth of ASW at 10 and 20 K. The ice cracking is accompanied by a sudden decrease in reflectance, in agreement with \citep{Bu2016a}. Although during ice growth at 10-20 K the cracking occurs spontaneously and suddenly, our experiments suggest a gradual structural change as the temperature is increased. 

The difference between the ice warmed up when deposited at 30~K and that at 110~K can be explained as follow. At 110 K, the ice is much more stable than at 30 K, and a very large critical thickness is needed for the ice to crack. Consequently, cracking does not happen in our experiment, and therefore the ice is more transparent. On the contrary, at 30 K, the ice is unstable and cracks fairly easily, and forms some gaps with a length scale comparable to or larger than the wavelength. Such a difference of behavior with the deposition temperature likely explains the decrease of extinction coefficient at a deposition temperature between 30 and 110 K.

While we are aware that ice porosity changes with different deposition temperatures and that when ice is heated, pores coalesce and become larger \citep{Stevenson1999, Bossa2014, He2019asw}, the dimensions of the pores are much smaller than the wavelength \citep{Raut2007, He2019asw, cazaux2015} and therefore pores do not contribute appreciably to the extinction. However, \citet{Bu2016b} showed that cracking of ASW films is caused by the intrinsic stress resulting from the film porosity, so a link between porosity (at the nanoscale) and cracking (and the micron-scale) does exist. It has been previously demonstrated by laboratory measurements \citep{naranen2004,shepard2007} and modeling \citep{hapke2008} that the scattering coefficient should increase with porosity. The difference in ice property between deposited ice and warmed ice is also demonstrated in prior experimental studies. For example, \citet{Hessinger1996} studied the internal friction (a mechanical property) of vapor-deposited ice and found that the ice deposited at low temperature and annealed to 160 K can have an internal friction value that is over one order of magnitude larger than that of ice deposited at 160 K. One corollary from the above finding is that the structure of ASW depends not only on the temperature the ice is annealed to but also on the temperature at which the ice is grown.

\section{Astrophysical implications}
\label{sec:astro_new}
The current study presents UV-Vis water refractive index and extinction/scattering coefficients at different temperatures. These data are important to interpret astronomical observations in our Solar System, where icy moons covered by water ice are found to have a very high albedo \citep{Cruikshank1980}, i.e., the ratio between the incident and reflected flux on a surface is typically close to one. In addition, when the effects of the temperature gradient found in these moons comes into place, temperature-dependent refractive indices are needed to avoid misleading conclusions.

The brightness of the diffuse reflectance depends on the competition between the scattering and the absorption. Our experimental results show that the extinction of the pure water ice, which is dominated by scattering, varies widely depending on the thermal history of the ice. For instance, if ice is grown by vapor condensation at a surface temperature around 110 K, the scattering coefficient is very small (see Fig.~\ref{fig:TPD_compare}). The existence of a small fraction of contaminants would reduce the brightness of the diffuse reflection. On the other hand, if the ice is grown by vapor condensation at a much lower surface temperature and subsequently warmed up, the scattering coefficient becomes much higher. Most of the light would be backscattered within a short distance and the absorbance by contaminants plays less of a role. In such a scenario, the ice would appear brighter. This is particularly relevant to icy moons in the Jupiter system such as Europa and Ganymede where some fraction of the water ice is thought to be in the amorphous form \citep{Clark2013}. Due to the diurnal or seasonal cycle, water evaporation and condensation are possible. In models of these icy moons in the UV-Vis range, one has to take into account the surface temperature when condensation takes place as well as the thermal history of the ice to accurately calculate the reflectance from these ice surfaces.

Among the icy moons in the Solar System, Iapetus - the third-largest natural satellite of Saturn - shows a color dichotomy in the visible part of the electromagnetic spectrum. As a consequence, the reflectance in the dark and bright regions is different. As observed by the Visible and Infrared Mapping Spectrometer (VIMS) instrument onboard the Cassini spacecraft the reflectance at 500~nm varies from 0.4 in the bright region to lower than 0.1 in the dark region \citep[][]{Clark2012}. The consensual explanation for this bimodal color distribution is the chemical composition of the surface. As explained by \citet{Clark2012}, while the bright region is by far dominated by H$_2$O ice, the dark area is composed of frozen water mixed with metallic iron, iron oxide (hematite), CO$_2$, traces of complex organic molecules, and yet unidentified materials. This conclusion was derived from Hapke models of the reflectance spectra of the Iapetus moon. \citet{Clark2012} observed that with the increase of the composition of nano-iron in the mixture with water ice, the reflectance decreases at wavelengths shortwards of 3~$\mu$m. Although iron compounds are indicated to be the cause of the reflectance decreasing, their origin is less understood \citep[][]{Ciarniello2019}. Possible sources of iron are attributed to meteoritic debris falling into Saturn's system. Hematite would be subsequently formed when the iron is oxidized closer to Saturn by oxygen sources in the atmosphere of Saturn's rings \citep[e.g.,][]{FILACCHIONE2021123}. Based on this study, we suggest that in addition to the interpretation by the iron contaminants, one should also consider the thermal history of the water ice. If the ice is grown at a temperature close to 110 K, the scattering would be low even if it is later heated up to crystallization temperature. In this case, even a small concentration of contaminants darkens the ice significantly. 

The temperature map of a selected region of the Iapetus moon \citep[][]{Spencer2010} was measured directly by Cassini's Composite Infrared Spectrometer (CIRS) using the short-wavelength detector (FP3; 600$-$1100~cm$^{-1}$). The area covered by CIRS shows that the temperatures at the brightest and darkest regions are around 100~K and 130~K, respectively. Despite this finding, the effect of the water ice temperature on the reflectance curves of Iapetus have not been considered yet. Based on the findings presented here, upcoming work will investigate whether and how the temperature-dependent refractive index may explain the color dichotomy found in Iapetus. This work will also take into account other celestial bodies in the outer Solar System.

Beyond the Solar System, in the ISM, the data presented in this paper is needed to properly interpret scattering effects toward young stellar objects and background stars. Usually, the experimental data used to model those objects are limited to the range between 2.0 and 20~$\mu$m \citep[e.g.,][]{Pontoppidan2005, rocha2015}, and outside of this spectral interval, the extinction efficiency relies on approaches of extrapolations. With the new temperature-dependent refractive index, this problem can be partially solved for the UV-vis spectral range.

\section{Conclusions}
The main conclusions of this study are summarized as follows:
\begin{itemize}
  \item A new method that combines HeNe and UV-Vis broadband interferometry is shown to effectively measure the refractive index and extinction coefficient of vapor-deposited water ice in the UV-Vis range. This method does not rely on the separate measurement of ice density or thickness and covers the UV-Vis range in one run.
  \item The refractive index of vapor-deposited water ice is not a monotonical function of the deposition temperature as previously reported. The value at 120 K is lower than that at 110 or 130 K. At this temperature, a large increase in the extinction coefficient has been found.
  \item The Lorentz-Lorenz relation does not apply for all temperatures of vapor-deposited water ice.
  \item The method introduced here also allows to derive the UV-Vis extinction coefficient for water ices deposited at different temperatures that is found to be non-negligible, contrary to previous studies presented in the earlier literature.
  \item Between a deposition temperature of 110 and 130 K, the extinction coefficient of vapor-deposited water ice at 500 nm increases by a factor of 8.5. This finding, combined with the lower refractive index value at 120 K, suggests a significant change in water structure between a deposition at 110 K and 130 K.
  \item The structure of vapor-deposited water ice depends not only on the temperature it is annealed to but also on the temperature it is deposited.
  \item The temperature dependence of the extinction coefficient in the UV-Vis range holds the potential to reveal the thermal history of water ice in the outer Solar System. 
  \item A good knowledge of the temperature-dependent refractive index will help in explaining astronomical observations including those of icy moons and ices in the ISM. While this data is entirely suitable to probe interstellar ices in cold molecular cores and high-temperature environments, such as the snowline regions in protoplanetary disks, the crystalline form ($>$120~K) is useful to interpret observational data in the outer Solar System.
  
\end{itemize}

\section{Acknowledgements}

This work has been financially supported by NWO (Netherlands Organisation for Scientific Research) through DANII, the Dutch Astrochemistry Network. WRMR thanks the financial support of the Leiden Observatory. We acknowledge the initial work by Vincent Kofman (see Kofman et al. 2019) that triggered the study presented here. We also thank Aditya M. Arabhavi, Stephanie Cazaux, Jordy Bouwman, Will Grundy, and Amanda Hendrix for insightful discussions.

\bibliographystyle{aasjournal}
\bibliography{ref} 

\end{document}